\documentclass[
  journal=pasa,
  manuscript=research-paper, 
  year=2025,
  volume=YY,
]{cup-journal}

\usepackage{amsmath}
\usepackage{amssymb,microtype,siunitx,booktabs}
\usepackage{amsmath,amssymb}
\usepackage{orcidlink}
\usepackage{caption}
\usepackage{aas_macros}
\usepackage{rotating}
\usepackage{tabularx,xltabular,supertabular}
\usepackage{multicol,longtable}
\usepackage{setspace,hyperref}
\hypersetup{colorlinks=true,
            linkcolor=blue,
            urlcolor=blue,
            linktoc=all,
            citecolor=blue}
            
\title{EMU/GAMA: A new approach to characterising radio luminosity functions}
\author{J. Prathap\textsuperscript{\orcidlink{0009-0004-0251-2672}}}
\affiliation{School of Mathematical and Physical Sciences, Macquarie University, Sydney, NSW 2109, Australia}
\alsoaffiliation{Macquarie University Astrophysics and Space Technologies Research Centre, Sydney, NSW 2109, Australia}
\email[J. Prathap]{jahangprathap12@gmail.com}

\author{A. M. Hopkins\textsuperscript{\orcidlink{0000-0002-6097-2747}}}
\affiliation{School of Mathematical and Physical Sciences, Macquarie University, Sydney, NSW 2109, Australia}
\alsoaffiliation{Macquarie University Astrophysics and Space Technologies Research Centre, Sydney, NSW 2109, Australia}

\author{J. Afonso\textsuperscript{\orcidlink{0000-0002-9149-2973}}}
\affiliation{Instituto de Astrofísica e Ciências do Espaço, Universidade de Lisboa, OAL, Tapada da Ajuda, PT1349-018 Lisbon, Portugal}
\alsoaffiliation{Departamento de F\'isica, Faculdade de Ci\^encias, Universidade de Lisboa, Edif\'ecio C8, Campo Grande, PT1749-016 Lisbon}

\author{M. Bilicki\textsuperscript{\orcidlink{0000-0002-3910-5809}}}
\affiliation{Centre for Theoretical Physics, Polish Academy of Sciences, Al. Lotnik\'ow 32/46, 02-668 Warsaw, Poland}

\author{M. Cowley\textsuperscript{\orcidlink{0000-0002-4653-8637}}}
\affiliation{School of Chemistry and Physics, Queensland University of Technology, Brisbane, QLD, Australia}
\alsoaffiliation{Centre for Astrophysics, University of Southern Queensland, West Street, Toowoomba, QLD 4350, Australia}

\author{S. M. Croom\textsuperscript{\orcidlink{0000-0003-2880-9197}}}
\affiliation{Sydney Institute for Astronomy (SIfA), School of Physics, A28, The University of Sydney, NSW 2006, Australia}

\author{Y. Gordon\textsuperscript{\orcidlink{0000-0003-1432-253X}}}
\affiliation{Physics Department, 2320 Chamberlin Hall, University of Wisconsin-Madison, 1150 University Avenue Madison, WI 53706-1390, USA}

\author{S. Phillipps\textsuperscript{\orcidlink{0000-0001-5991-3486}}}
\affiliation{Astrophysics Group, School of Physics, University of Bristol, Tyndall Avenue, Bristol BS8 1TL, UK}

\author{E. M. Sadler\textsuperscript{\orcidlink{0000-0002-1136-2555}}}
\affiliation{Sydney Institute for Astronomy (SIfA), School of Physics, A28, The University of Sydney, NSW 2006, Australia}
\alsoaffiliation{CSIRO Space and Astronomy, PO Box 76, Epping NSW 1710, Australia }

\author{S. S. Shabala\textsuperscript{\orcidlink{0000-0001-5064-0493
}}}
\affiliation{School of Natural Sciences, University of Tasmania, Private Bag 37, Hobart, Tasmania 7001, Australia}

\author{U. T. Ahmed\textsuperscript{\orcidlink{0000-0002-0309-1599
}}}
\affiliation{Australian Astronomical Optics, Macquarie University, 105 Delhi Rd, North Ryde, NSW 2113, Australia}
\alsoaffiliation{Macquarie University Astrophysics and Space Technologies Research Centre, Sydney, NSW 2109, Australia}
\alsoaffiliation{Centre for Astrophysics, University of Southern Queensland, 37 Sinnathamby Boulevard, Springfield Central, QLD 4300, Australia}

\author{S. Amarantidis\textsuperscript{\orcidlink{0000-0001-7948-5714}}}
\affiliation{Institut de Radioastronomie Millim\'etrique (IRAM), Avenida Divina Pastora 7, Local 20, E-18012, Granada, Spain}

\author{M. J. I. Brown\textsuperscript{\orcidlink{0000-0002-1207-9137}}}
\affiliation{School of Physics, Monash University, Clayton, VIC 3800, Australia}

\author{R. Carvajal\textsuperscript{\orcidlink{0000-0002-0545-1113}}}
\affiliation{Instituto de Astrofísica e Ciências do Espaço, Universidade de Lisboa, OAL, Tapada da Ajuda, PT1349-018 Lisbon, Portugal}
\alsoaffiliation{Departamento de F\'isica, Faculdade de Ci\^encias, Universidade de Lisboa, Edif\'ecio C8, Campo Grande, PT1749-016 Lisbon}

\author{D. Leahy\textsuperscript{\orcidlink{0000-0002-4814-958X}}}
\affiliation{Department of Physics and Astronomy, University of Calgary, Calgary, AB T2N 1N4, Canada}

\author{J. R. Marvil\textsuperscript{\orcidlink{0000-0003-1111-8066}}}
\affiliation{National Radio Astronomy Observatory, P.O. Box O, Socorro, NM 87801, USA}

\author{T. Mukherjee\textsuperscript{\orcidlink{0009-0004-7639-869X}}}
\affiliation{School of Mathematical and Physical Sciences, Macquarie University, Sydney, NSW 2109, Australia}
\alsoaffiliation{Macquarie University Astrophysics and Space Technologies Research Centre, Sydney, NSW 2109, Australia}

\author{J. Willingham\textsuperscript{\orcidlink{0009-0001-5653-9481}}}
\affiliation{School of Mathematical and Physical Sciences, Macquarie University, Sydney, NSW 2109, Australia}
\alsoaffiliation{Macquarie University Astrophysics and Space Technologies Research Centre, Sydney, NSW 2109, Australia}

\author{T. Zafar\textsuperscript{\orcidlink{0000-0003-3935-7018}}}
\affiliation{School of Mathematical and Physical Sciences, Macquarie University, Sydney, NSW 2109, Australia}
\alsoaffiliation{Macquarie University Astrophysics and Space Technologies Research Centre, Sydney, NSW 2109, Australia}

\received {dd Mmm YYYY}
\revised  {dd Mmm YYYY}
\accepted {dd Mmm YYYY}
\published{dd Mmm YYYY}

\keywords{radio luminosity functions, radio AGN, radio surveys, statistical techniques} 
\begin{document}
\begin{abstract}
This study characterises the radio luminosity functions (RLFs) for SFGs and AGN using statistical redshift estimation in the absence of comprehensive spectroscopic data. Sensitive radio surveys over large areas detect many sources with faint optical and infrared counterparts, for which redshifts and spectra are unavailable. This challenges our attempt to understand the population of radio sources. Statistical tools are often used to model parameters (such as redshift) as an alternative to observational data. Using the data from GAMA G23 and EMU early science observations, we explore simple statistical techniques to estimate the redshifts in order to measure the RLFs of the G23 radio sources as a whole and for SFGs and AGN separately. Redshifts and AGN/SFG classifications are assigned statistically for those radio sources without spectroscopic data. The calculated RLFs are compared with existing studies, and the results suggest that the RLFs match remarkably well for low redshift galaxies with an optical counterpart. We use a more realistic high redshift distribution to model the redshifts of (most likely) high redshift radio sources and find that the LFs from our approach match well with measured LFs. We also look at strategies to compare the RLFs of radio sources without an optical counterpart to existing studies.
\end{abstract}

\section{Introduction}\label{sec:intro}
Mapping the distribution of galaxies is fundamental in understanding the large scale matter distribution of the Universe, for which luminosity functions (LFs) have been instrumental throughout. The galaxy LF is the comoving source density with luminosity and is a statistical quantity used to understand the galaxy distribution. Redshift evolution of the LFs quantifies the cosmic evolution of the statistical properties of galaxy population \citep[e.g.,][]{1977AJ.....82..861F,2015MNRAS.447..875G,2017A&A...602A...6S,2017A&A...602A...5N,peters2019luminosity}. The population of star forming galaxies (SFGs) and active galactic nuclei (AGN) exhibit different characteristics as a function of luminosity and redshift. Consequently, separate analyses are important in disentangling the cosmic evolution of these populations \citep[e.g.,][]{2007MNRAS.375..931M,2017A&A...602A...5N,2017A&A...602A...6S}.

To characterise the evolution of these distinct populations, advancements in multiwavelength astronomy have been critical, enabling LFs to be measured across the entire electromagnetic spectrum. Ultraviolet (UV) and optical LFs, owing to a large number of space and ground-based surveys, benefit from having large samples, which results in a better constrained measurement \citep[e.g.,][]{2004MNRAS.349.1397C,2015ApJ...803...34B}. However, both need to account for dust obscuration. The infrared (IR) regime can trace the obscured stellar/AGN light but is not sensitive to emission that freely escapes from the galaxy. Radio emission, on the other hand, is not affected by dust and, hence, is a valuable alternative as a probe of the galaxy population independent of dust obscuration \citep[e.g.,][]{2007MNRAS.375..931M,2017A&A...602A...5N,2017A&A...602A...6S}. Luminous radio emission typically arises from AGN, while low-luminosity radio emission is typically from star formation in nearby galaxies \citep{1966MNRAS.133..421L}. The faint radio source population was originally probed through single pointing radio surveys \citep[e.g.,][]{1984A&AS...58....1W,1985ApJ...289..494W}. In the past few decades, this has been significantly expanded through the advent of deep and wide surveys \citep{2002MNRAS.329..227S,2007MNRAS.375..931M,2016MNRAS.460....2P,peters2019luminosity}.

Because of cosmic variance, many factors go into calculating LFs when accounting for the Universe as a whole. No single survey can detect all the sources in a given sky area to provide a complete, unbiased sample. The detection probability of an object is affected by various factors, such as the survey sensitivity, the input catalogue of the survey, its target and spectroscopic completeness, area and depth, and the observing conditions. The intrinsic properties of a source, such as its surface brightness and size, can also influence whether it gets detected. The luminosity of a source, together with these factors, governs its likelihood of detection. Consequently, low luminosity sources are typically the fewest in any magnitude limited survey \citep[e.g.,][]{2012AJ....143..102G,2015MNRAS.447..875G}. Although they do not contribute much to the overall luminosity in proportion, faint galaxies greatly outnumber their bright counterparts. Standard methods are often used to account for these biases and incompleteness (\citealt{1968ApJ...151..393S}, also see \S\,\ref{sec:v_max}).

Once the LF is corrected for sample incompleteness and any biases, it is typical to represent the function using an analytical form. A common representation is that of \citet{1976ApJ...203..297S}:
\begin{equation}
    \phi(L)~dL=\frac{\phi_*}{L_*}\left(\frac{L}{L_*}\right)^a\,e^{-L/L_*}\,dL,
    \label{eq:schechter}
\end{equation}
where $\phi_*$ is the normalisation factor that defines the overall density of galaxies, $L_*$ is the characteristic galaxy luminosity (an $L_*$ galaxy is comparable to the Milky Way, whereas a galaxy of $L<0.1~L_*$ is a dwarf), and $a$ defines the faint end slope. This functional form behaves as a power law with slope $a$ for luminosities $L<L_*$ and as an exponential for $L>L_*$. While the Schechter function provides a useful framework for representing LFs across various wavelengths, RLFs offer a unique advantage by being insensitive to dust obscuration.
\subsection{RLFs}\label{sec:radio_lf}
The RLF represents the source density of radio galaxies per unit comoving volume distributed in bins of radio luminosity. The radio source population has been extensively explored using RLFs \citep[e.g.,][]{2007MNRAS.381..211S,2007MNRAS.375..931M,2012MNRAS.421.1569B,2017A&A...602A...5N,2017A&A...602A...6S,2018A&A...620A.192C,2019ApJ...872..148C,2024MNRAS.532.1504J,2024A&A...685A..79W,2025MNRAS.536L..32M}. In addition, RLFs have also been used in probing radio source clustering \citep{1998MNRAS.300..257M,2003A&A...405...53O}. The population of galaxies with radio emission is a mix of sources powered by AGN and SFGs \citep[e.g.,][]{2016MNRAS.457..629C}. Classical radio galaxies, typically ellipticals, are either compact or multicomponent \citep{2008MNRAS.388..625S,2019A&A...622A..12H}, and are driven by AGN \citep{1989ApJ...338...13C,1992ARA&A..30..575C,1999MNRAS.308...45M,2009AJ....137.4450M}. This is usually the case for radio galaxies with $\rm L_{1.4\,\rm{GHz}}\geq10^{23}\,W\,Hz^{-1}$ as evident from local 1.4 GHz RLFs \citep{2002AJ....124..675C,2007MNRAS.375..931M,2012MNRAS.421.1569B}. Below this threshold, an increasing number of radio-emitting SFGs are found \citep{1989ApJ...338...13C,2008MNRAS.386.1695S}.

For the current generation of deep radio surveys, a low redshift sample contains a significant number of star formers with AGN dominating at higher redshifts \citep{2001MNRAS.322..536W}. Consequently, the shape and evolution of the RLF with redshift constrain the nature of activity in massive galaxies and their evolution. 
Just as UV and optical LFs are commonly parameterised using the Schechter function (Eq. \ref{eq:schechter}), IR and radio LFs are commonly parameterised using the Saunders function \citep{1990MNRAS.242..318S,2007MNRAS.375..931M}:
\begin{equation}
    \phi(L)dL=\frac{\phi_*}{L_*}\left(\frac{L}{L_*}\right)^a \exp \left[-\left(\frac{\log(1+L/L_*)}{\sqrt{2}\sigma}\right)^2\right]~dL.
    \label{eq:saunders}
\end{equation}
It behaves in the same fashion as the Schechter function for $L<L_*$ and as a Gaussian (rather than exponential) in $\log\,L/L_*$ with a width $\sigma$ for $L>L_*$. Radio AGN, on the other hand, are commonly parameterised using a two power-law function \citep[][also see \S\,\ref{sec:lf_results}]{1990MNRAS.247...19D,2001AJ....121.2381B}.

\citet{1966MNRAS.133..421L}, one of the first to attempt to determine the evolution of radio source population, found that the models where the most powerful radio sources undergo greater comoving space density evolution than less powerful sources best fitted the data. Evolution is generally incorporated into the estimated RLF by two models: pure density evolution (PDE) and pure luminosity evolution (PLE) (e.g., \citealt{1996MNRAS.281..953P,2007A&A...461...39F,2017A&A...602A...5N}, also see luminosity dependent density evolution (LDDE), \citealt{1983ApJ...269..352S}). 

In PDE, the number density of galaxies is free to evolve with redshift, and their luminosities are presumed to remain the same. Physically, this can be achieved through merger events (not generally true for classical radio sources). If $\phi_0(L)$ is the local RLF, PDE can be incorporated into it as:
\begin{equation}
    \phi_z(L,z)  = (1+z)^{k_D}\,\phi_0(L),
    \label{eq:pde}
\end{equation}
where $k_D$ is the free parameter that determines the rate of evolution \citep{1993MNRAS.263..123R,2004ApJ...615..209H,peters2019luminosity}.

PLE, on the other hand, assumes no evolution in the galaxy demography. Rather, it includes changes in galaxy luminosities with redshift. Galaxies become fainter due to star formation quenching, such as through AGN feedback or stellar winds. A sudden influx of cold gas or AGN feedback can boost star formation and increase the luminosity of the galaxy as a whole. The redshift dependence of the LF is governed by the evolution in luminosity parameterised as:
\begin{equation}
    \phi_z(L,z)=\phi_0 \left(\frac{L}{(1+z)^{k_L}}\right),
    \label{eq:ple}
\end{equation}
where $k_L$ is the free parameter that determines the rate of evolution based on the changes in luminosity, as opposed to density in PDE.

\subsection{The need for redshift}\label{sec:intro_z}
Spectroscopic redshifts of sources are crucial in measuring the LF and in any analysis requiring AGN classification \citep[e.g.,][]{2002AJ....124..675C,2002MNRAS.329..227S,2005MNRAS.362....9B,2007MNRAS.375..931M}. In the context of radio surveys, however, the tens of millions of radio sources being discovered are rapidly outstripping the capabilities of existing spectroscopic facilities \citep[e.g.,][]{2023PASA...40...39L}.

Photometric template fitting \citep{1957AJ.....62....6B,1986ApJ...303..154L} has been extensively used in the absence of spectroscopic redshifts, often approaching comparable accuracy for typical galaxy populations, especially at lower redshifts \citep[$z\lessapprox1$, see e.g.,][]{2009ApJ...690.1236I,2016ApJ...830...51S,2024AJ....168..233L,2024A&A...691A.308S}. With the new generation of radio surveys, (e.g., Evolutionary Map of the Universe (EMU), \citealt{2011PASA...28..215N}, LOw Frequency ARray (LOFAR) Two-metre Sky Survey (LoTSS),  \citealt{2017A&A...598A.104S}, GaLactic and Extragalactic All-sky Murchison Widefield Array survey eXtended (GLEAM-X), \citealt{2022PASA...39...35H}) the required photometric coverage is now the limiting factor for a large number of radio sources. The difficulty in disentangling star forming and AGN emission as well as the lack of specialised templates modelling AGN further complicates the template fitting approach for radio sources \citep{2019NatAs...3..212S,2023PASA...40...39L}.

When photometric or spectroscopic redshifts are not feasible for all sources, one can leverage statistical techniques to model redshifts of radio sources informed by available data from similar sources. This is done by looking for correlations between observed properties of the radio source, such as its redshift and flux density, with some underlying assumptions and then extending the relation to the source with unknown redshift.

Similar to its application in various other fields, machine learning (ML) techniques have been used in astronomy for redshift estimation \citep{2023RSOS...1021454S,2023arXiv230400512W}. For instance, algorithms such as k-nearest neighbours (kNN; \citealt{1053964}) in \citet{2017MNRAS.465.1959C} and \citet{2023PASA...40...39L}, random forest (RF; \citealt{598994}) in \citet{2017MNRAS.465.1959C} and \citet[][see also Carvajal et al., in prep.]{2021Galax...9...86C}, neural networks (NNs) in \citet{2022MNRAS.512.2099C} and \citet{2023PASA...40...39L}, and Gaussian processes (GPs) in \citet{2018MNRAS.473.2655D,2018MNRAS.477.5177D,2021A&A...648A...4D} and \citet{2023PASA...40...39L} demonstrate the capability of ML techniques in redshift estimation. \citet{2022MNRAS.509.5467K} used a different approach in which the evolving galaxy LF was measured with a clustering-based redshift inference technique using spatial cross-correlation statistics. 

In this work, we explore simple statistical techniques that draw on minimal available information in order to derive RLFs, anticipating future large datasets where the large numbers of deep multiwavelength photometric measurements needed by other techniques may not be available. Here we implement such an approach to model the redshift distributions of radio sources from EMU early science data \citep{2022MNRAS.512.6104G} and derive their LFs. This data is cross-matched with the Galaxy and Mass Assembly (GAMA; \citealt{2011MNRAS.413..971D,2020MNRAS.496.3235B}) survey data release 4 (DR4; \citealt{2022MNRAS.513..439D}) for spectroscopic redshifts and $r$-band apparent magnitudes ($m_r$). Radio sources with spectroscopic counterparts serve as priors for redshift estimation, while the $m_r$, where available, supplements this estimation by helping to constrain redshifts for sources lacking spectroscopic data.

The structure of the paper is as follows: The data are described in \S\,\ref{sec:lf_data}. \S\,\ref{sec:z_estimation} details the statistical approach in estimating the redshifts of radio sources with and without optical counterparts. The radio sources with spectroscopic redshifts are classified into SFGs and AGN in \S\,\ref{sec:agn_diagnostics}. This section also extends the classification to sources without spectroscopic redshifts. The RLFs are derived in \S\,\ref{sec:rlf} and the initial results are described in \S\,\ref{sec:lf_results}. \S\,\ref{sec:lf_discussion} discusses the main findings. \S\,\ref{sec:conclusion} summarises the work and concludes the paper. Throughout the paper we assume the \citet{2016A&A...594A..13P} cosmology with $\rm H_0 = 67.8\,{km\,s^{-1}\,Mpc^{-1}}$, $\Omega_M = 0.308$, and  $\Omega_\Lambda = 0.692$. 
\section{Data}\label{sec:lf_data}
\subsection{GAMA}\label{sec:gama_data}
GAMA is a multiband imaging and spectroscopic survey undertaken at the Anglo Australian Telescope (AAT) using the two-degree field (2dF) fibre feed and the AAOmega spectrograph. The survey provides high quality spectroscopy covering five separate fields (G02, G09, G12, G15, and G23) with a limiting magnitude of $m_r<19.8$ mag ($m_i<19.2$ mag for the G23 field), collectively spanning an extensive area of 286 square degrees.

Four of the GAMA fields (G09, G12, G15, and G23) benefit from the overlap with the footprints of the European Southern Observatory (ESO) VST Kilo Degree Survey \citep[KiDS,][]{2015A&A...582A..62D}, ESO Visible and Infrared Survey Telescope for Astronomy (VISTA) Kilo-degree Infrared Galaxy Public Survey \citep[VIKING,][]{2013Msngr.154...32E}, Galaxy Evolution Explorer \citep[GALEX,][]{2005ApJ...619L...1M}, Wide-field Infrared Survey Explorer \citep[WISE;][]{2010AJ....140.1868W}, and Herschel \citep{2010A&A...518L...1P} surveys, ensuring extensive multiwavelength imaging and photometric coverage ranging from FUV to FIR. The GAMA photometric catalogue has a specific flux cut applied on the KiDS r-band magnitude \citep[$<$ 20 mag,][]{2020MNRAS.496.3235B}, whereas the underlying KiDS and VIKING data go much deeper \citep{2024A&A...686A.170W}. The analysis in this work thus can be extended to include current and upcoming deep photometric data \citep[e.g.,][]{2017arXiv170804058L}. 

The latest data release of GAMA, DR4 \citep{2022MNRAS.513..439D} combines photometric data from the aforementioned surveys together with the spectroscopy from the AAOmega spectrograph. We are interested in the G23 field ($\rm 338.1^o<RA<351.9^o$, $-35^o<\delta<-30^o$) since it overlaps with the EMU early science observation \citep{2022MNRAS.512.6104G}. 

In this paper, we use $m_r$ from the \verb|gkvScienceCatv02| table \citep{2020MNRAS.496.3235B} in the \verb|gkvInputCatv02| data management unit (DMU), with 458\,844 sources. The \verb|SpecLineSFRv05| DMU contains the \verb|GaussFitSimplev05| table (\citealt[][49\,623 sources]{2017MNRAS.465.2671G}) from which spectroscopic redshifts are retrieved. We use these two tables to perform multiple crossmatches with the radio sample defined in the next section.  
\subsection{EMU}\label{sec:emu_data}
The Australian Square Kilometre Array Pathfinder (ASKAP, \citealt{2007PASA...24..174J, 2016PASA...33...42M, 2021PASA...38....9H}), a precursor of the Square Kilometre Array (SKA), consists of 36\,\verb|x|\,12-m antennae each of which is equipped with a phased array feed (PAF) resulting in a field of view of $\rm 30\,deg^2$ and high survey speeds. ASKAP is designed to observe in the 800–1800 MHz frequency range with an instantaneous bandwidth of 288 MHz and offers rapid observations with good resolution and sensitivity.

EMU \citep{2011PASA...28..215N,2021PASA...38...46N} is an ongoing wide-field radio continuum survey conducted using ASKAP, providing a deep radio continuum map of the entire southern sky, extending up to $\delta=0^o$ at 943 MHz. The survey achieves a sensitivity of approximately $20\,\mu\rm\,Jy\,beam^{-1}$ and a resolution of 15" and is currently around 20\% complete in terms of observations. Upon completion, it is anticipated that EMU will detect and catalogue roughly 20 million galaxies, including SFGs up to $z\sim1$, powerful starbursts reaching even greater redshifts, and AGN to the edge of the visible Universe (Hopkins et al., submitted).

The multiwavelength FUV-FIR photometric coverage of the GAMA G23 region is augmented by radio data through the ASKAP commissioning \citep{2019PASA...36...24L} and later as part of the EMU early science program \citep{2022MNRAS.512.6104G}. In this paper, we use the data from the early science program that covers an area of 82.7 deg$^2$ and achieves a sensitivity of $\rm0.038\,m\,Jy\,beam^{-1}$ at a frequency of 887.5 MHz. The survey detected 55\,247 sources, out of which 39\,812 have signal-to-noise (S/N) ratios $\geq$ 5. The brightest radio galaxies are, in general,  depending on the angular resolution of the survey, multi component. In the \citet{2022MNRAS.512.6104G} catalogue, however, we have $<$ 1\% multi component sources, also their radio luminosities are of the order of single component sources. Because of this, we do not expect any potential biases due to the existence of multi component sources in the sample. We use the parameter \verb|Total_flux| from \citet{2022MNRAS.512.6104G} corresponding to the measured integrated flux from the radio source at 887.5 MHz. This radio sample is referred to as `EMU' throughout the paper.
\begin{figure}[h!]
    \centering
    \includegraphics[width=\linewidth]{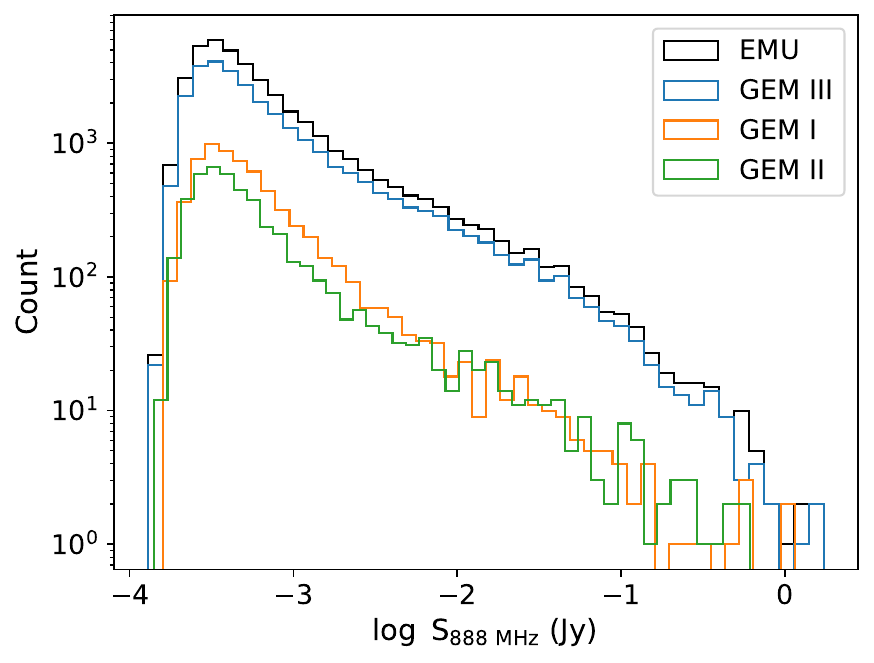}
    \caption{The 888 MHz radio flux density distributions of EMU (black), GEM\,III (blue), GEM\,I (orange), and GEM\,II (green) samples.}
    \label{fig:S_dist}
\end{figure}
\begin{figure}[h!]
    \centering
    \includegraphics[width=\linewidth]{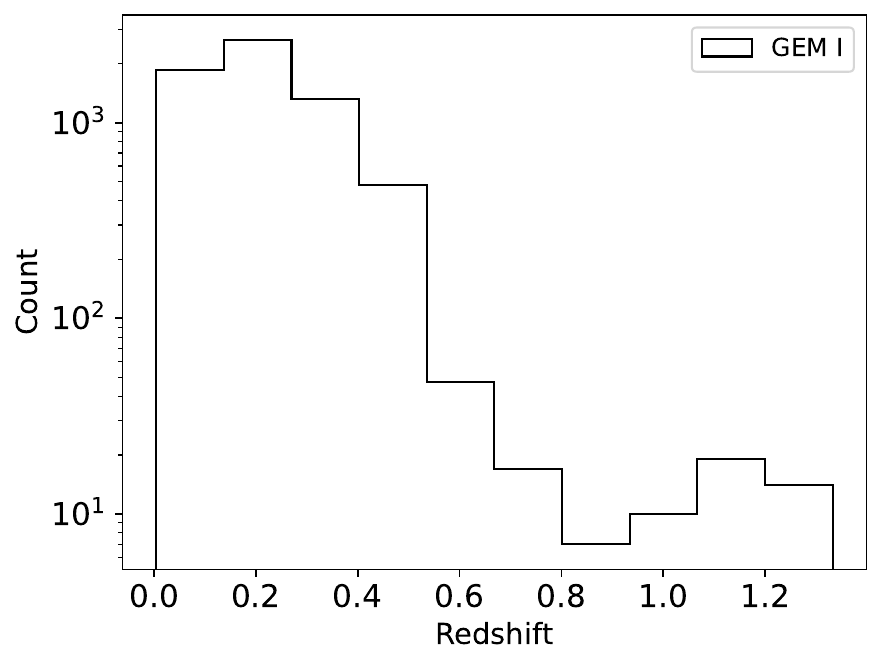}
    \caption{The redshift distribution of GEM\,I radio sources. The redshifts of these sources are the references for those without redshifts.}
    \label{fig:gemI_z}
\end{figure}
\subsection{Final Sample}\label{sec:final_sample}
To facilitate our analysis, we first implement two cross-matches between the radio and optical samples using Topcat \citep{2005ASPC..347...29T}. A cross-match between EMU and the \verb|gkvScienceCatv02| table with a cross-match radius of $5''$ \citep{2021PASA...38...46N,2024PASA...41...21A} results in 10\,991 radio sources with $m_r$. This table is then cross-matched with the \verb|GaussFitSimplev05| spectroscopic data, resulting in 6425 radio sources with both spectroscopic redshifts and $m_r$, and 4566 radio sources with only $m_r$.
\begin{table}[h!]
    \centering
    \begin{tabular}{c|c|c}
    \hline
         Sample & Number of sources & Optical data\\
         \hline
         GEM\,I & 6425 & $z$, $m_r$ \\
         GEM\,II & 4566 & $m_r$ \\
         GEM\,III & 28\,821 & $-$ \\
         \hline
         EMU & 39\,812 & $-$ \\
         \hline
    \end{tabular}
    \caption{The number of sources in different samples. GEM\,I sources have both spectroscopic redshifts ($z$), and $m_r$; GEM\,II has only $m_r$; GEM\,III has neither. EMU represents the entire radio sample.}
    \label{tab:lf_sample}
\end{table}
\begin{figure*}[h!]
    \centering
    \includegraphics[width=0.95\linewidth]{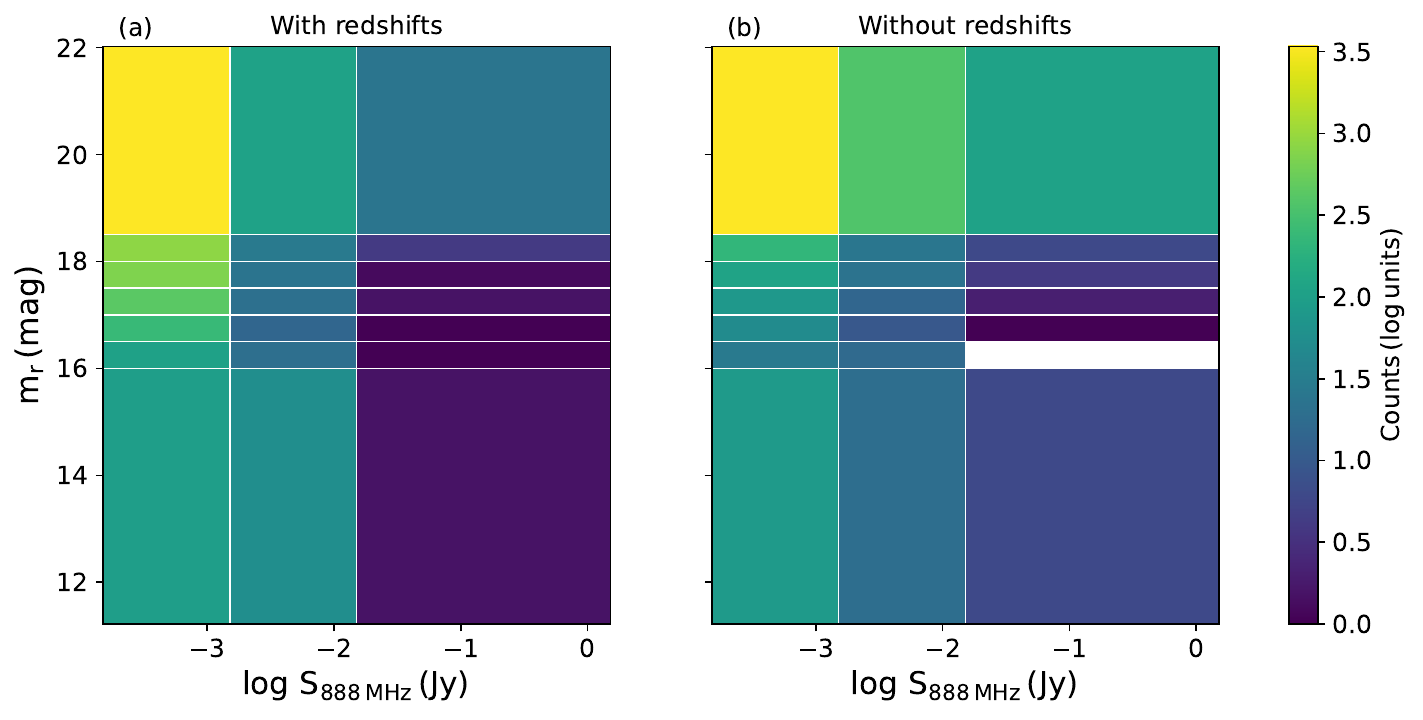}
    \caption{2D histograms of radio flux densities and r-band magnitudes, (a): GEM\,I, (b): GEM\,II. Each bin is colour-coded based on the number counts. In both panels, radio sources are preferentially distributed towards lower radio flux densities (below $\rm\log\,S_{888\,MHz}=-2.5\,Jy$) and fainter r-band magnitudes (above $m_r=16$). One white-coloured bin in (b) corresponds to zero sources in that bin.}
    \label{fig:r_z_2d}
\end{figure*}
\begin{figure*}[h!]
    \centering
    \includegraphics[width=0.95\linewidth]{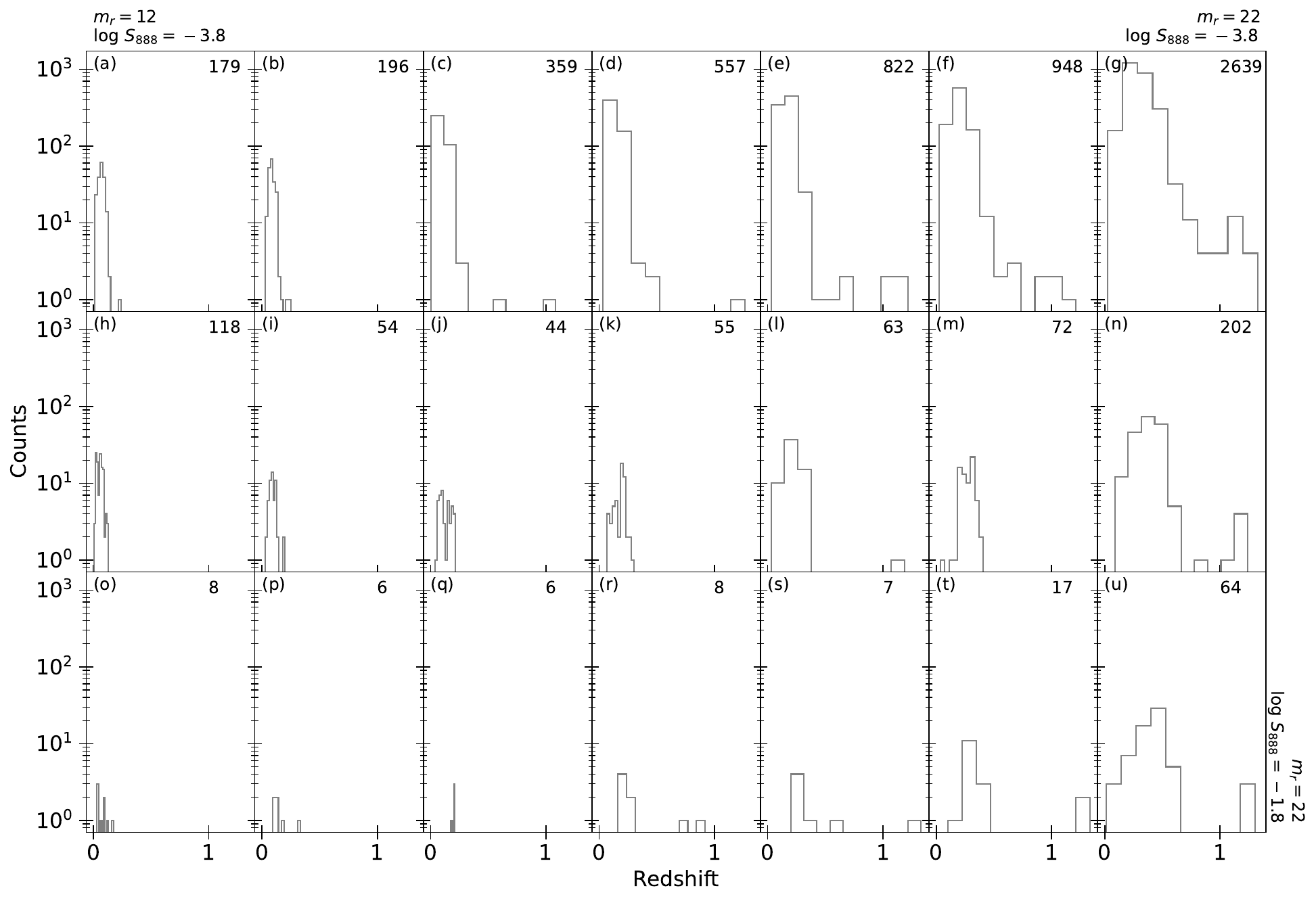}
    \caption{The redshift distribution of the GEM\,I sources corresponding to figure \ref{fig:r_z_2d}a. The x-axis shows the redshift, and the counts are shown in the y-axis. The number of sources in each bin is shown in the upper right corner. The $m_r$ and $S_{888}$ values shown outside the panel correspond to the respective bin positions in figure \ref{fig:r_z_2d}a.}
    \label{fig:z_panel}
\end{figure*}

Consequently, we have three sets of catalogues: $(i)$ radio sources with $m_r$ and spectroscopic redshifts (6425 sources), $(ii)$ radio sources with $m_r$ but not spectroscopic redshifts (4566 sources), and $(iii)$ radio sources with no optical counterparts (28\,821 sources). These samples are referred to as GEM\,I, GEM\,II, and GEM\,III, respectively (for GAMA-EMU, see Table \ref{tab:lf_sample}). GEM\,I facilitates modelling the redshifts for both GEM\,II and GEM\,III as detailed in \S\,\ref{sec:z_estimation}. GEM\,II has the advantage of having $m_r$, which further constrains the estimated redshifts. Figure \ref{fig:S_dist} illustrates the radio flux density distributions of these samples and Figure \ref{fig:gemI_z} shows the redshift distribution of GEM\,I sources.

As discussed in \S\,\ref{sec:intro}, survey completeness plays a crucial role in LF measurements. Figure 7 in \citet{2022MNRAS.512.6104G} shows the completeness of the EMU sample as a function of source flux density. The EMU sample is more than 95\% complete above a flux density of 1 mJy. Since we are estimating the redshifts of the entire EMU sample, the completeness values from \citet{2022MNRAS.512.6104G} can be applied directly to our calculations.
\section{Redshift estimation}\label{sec:z_estimation}
To accurately construct RLFs for AGN and SFGs, this study applies simple statistical techniques to model redshift distributions of radio sources. We start with the assumption that radio sources with similar radio flux densities and, when available, similar $m_r$, share comparable redshift distributions. We refine our assumptions in order to make them more realistic in later sections (see \S\,\ref{sec:z_redist} and \S\, \ref{sec:lf_discussion}). 

It is worth noting here that this work anticipates future large datasets where large numbers of multiwavelength counterparts may not be available. This is the reason for using an observed parameter like radio flux density to perform the analysis. It is important to gauge the potential value in analyses that draw only on the fewest observed parameters.

From Figure \ref{fig:gemI_z}, it is evident that most of the radio sources with spectroscopic redshifts lie at $z<0.5$, which is expected since the targets are selected from GAMA with a limiting magnitude of $m_r<19.8$ \citep[$m_i<19.2$;][]{2011MNRAS.413..971D,2022MNRAS.513..439D}. The following subsections describe our statistical redshift estimation procedure. GEM\,II and GEM\,III sources are discussed separately for clarity. 
\begin{figure}[h!]
    \centering
    \includegraphics[width=\linewidth]{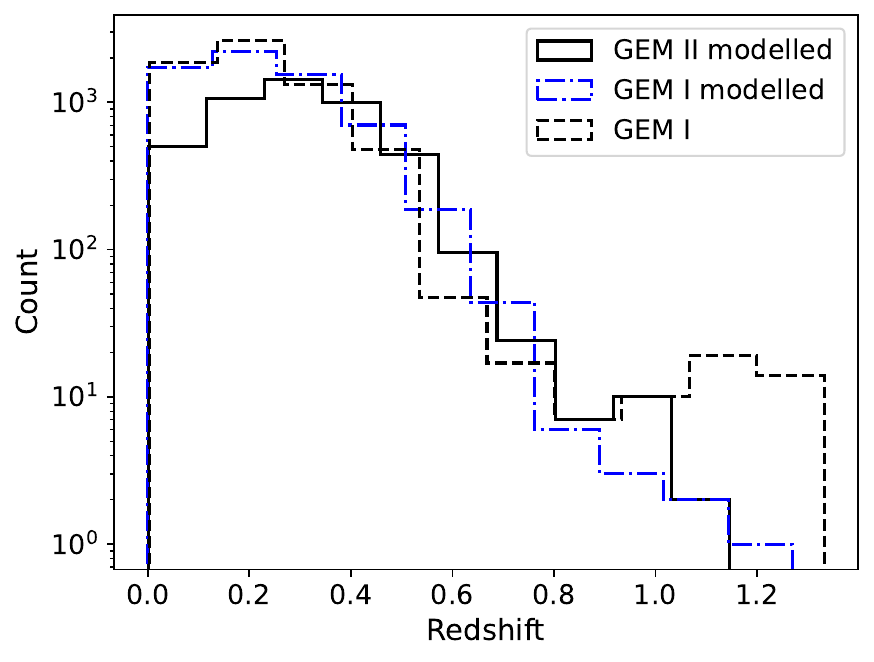}
    \caption{The modelled redshift distribution of the GEM\,II sample (solid line). The redshift distribution of the GEM\,I sample used for modelling is also shown (dashed line). The blue dot-dashed line shows the GEM I redshift distribution modelled using GEM I spectroscopic redshifts. The model reproduces the actual GEM I distribution, except for $z>1$, details of which are given in the text (\S\,\ref{sec:emu_r-band}). }
    \label{fig:2d_z}
\end{figure}
\subsection{GEM II: Sources with r-band magnitude}\label{sec:emu_r-band}
Figure \ref{fig:r_z_2d} shows the radio flux density and $m_r$ distributions, as 2D histograms, of GEM I (left panel) and GEM II (right panel) samples. The bins are colour coded according to the number counts as shown in the colour bar. 

Each bin in Figure \ref{fig:r_z_2d}a has a counterpart in Figure \ref{fig:r_z_2d}b and a corresponding redshift distribution. The bins are non-uniformly spaced so that at least six sources exist in any given bin to facilitate the fitting procedure described below to model the redshift distribution. Figure \ref{fig:z_panel} shows the redshift distribution and source counts corresponding to each bin in Figure \ref{fig:r_z_2d}a. In Figure \ref{fig:z_panel}, $m_r$ increases across the x-axis (redshift) from the origin, 
and radio flux density decreases along the y-axis (count) from the origin.

We use the \verb|fitter|\footnote{\url{https://github.com/cokelaer/fitter}} package for fitting normal distributions to each of the bins in Figure \ref{fig:z_panel}. The package uses the \verb|fit| method of \verb|SciPy| to extract the parameters of the distribution that best describes the data. The redshift distributions in each bin were tested for their normality using the quantile-quantile (Q-Q) plot. In most cases, the data are consistent with a Gaussian shape. This justifies the choice of a normal distribution in the fitting procedure. If, however, instead of choosing to fit a Gaussian, we assign redshifts randomly from the GEM\,I redshift distribution in a given bin, it yields comparable results for the derived RLFs. We make sure to avoid any negative redshifts given out by the model.

For a given bin, numbers are drawn randomly from the resulting fit using inbuilt random number generators in \verb|Python| and are assigned as the redshift to each source in that bin of GEM\,II (those sources without spectroscopic redshifts). This process is repeated for all the bins, and redshifts are generated for the entire GEM\,II sample. Figure \ref{fig:2d_z} shows how the resulting estimated GEM\,II redshifts are distributed (solid histogram). 

We tested our statistical approach using the GEM\,I redshift distribution, shown in Figure \ref{fig:2d_z} (blue dot-dashed histogram), which it accurately reproduces, except for $z\geq1$. This difference for $z\geq1$ is clear when compared to the GEM I spectroscopic redshift distribution (black dashed histogram). The underestimate here is a result of our simple single Gaussian model for the redshift distributions, which is limited when the reference set has only a few objects. This could be improved (and to include the radio sources with $z>1$) using multiple Gaussians or Gaussian mixture models (GMMs). When these methods are applied to GEM\,II and GEM\,III samples, the modelled redshift distributions reliably reproduce that of GEM\,I, as expected. This is not actually desirable, however, since we do not want the GEM\,I redshift distribution to be duplicated in fine detail. The choice to adopt single Gaussians in modelling the redshifts has the effect of smoothing over such fine features, that would correspond to large scale structures or potentially rare sources at $z\geq1$. Smoothing over these features ensures that the inferred redshift distribution follows the broad overall trend, rather than reproducing fine features that are not appropriate for the entire population.

\subsection{GEM III: Sources with no r-band magnitude}\label{sec:emu_no_r-band}
\begin{figure}[h!]
    \centering
    \includegraphics[width=\linewidth]{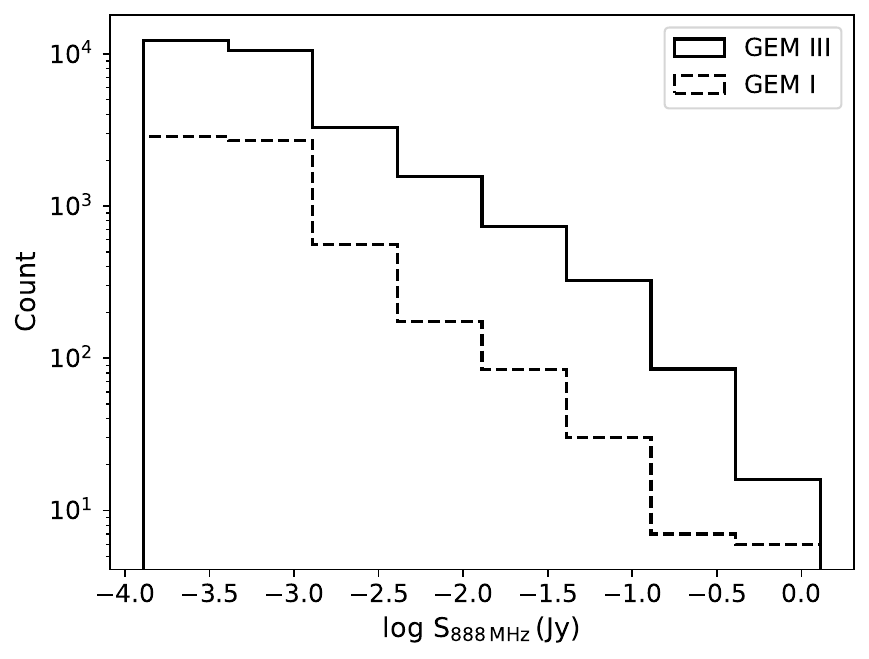}
    \caption{The 888 MHz radio flux density distributions of GEM\,I (dashed line) and GEM\,III (solid line) samples. The number counts increase to fainter radio flux densities in the same way as GEM\,II. However, a noticeable number of GEM\,III sources exist at higher radio flux densities.}
    \label{fig:S_1d}
\end{figure}

Figure \ref{fig:S_1d} shows the radio flux density distributions of GEM\,I (dashed histogram) and GEM\,III (solid histogram) samples. GEM\,III is significantly larger in size than GEM\,I, reflecting the fact that the full sample of radio sources far outstrips those with spectroscopic counterparts. Most of these radio sources are expected to reside at a redshift beyond the maximum limit probed by GAMA. So, we treat any results with caution while performing this initial analysis using these predictions for GEM\,III.
\begin{figure*}[h!]
    \centering
    \includegraphics[width=\linewidth]{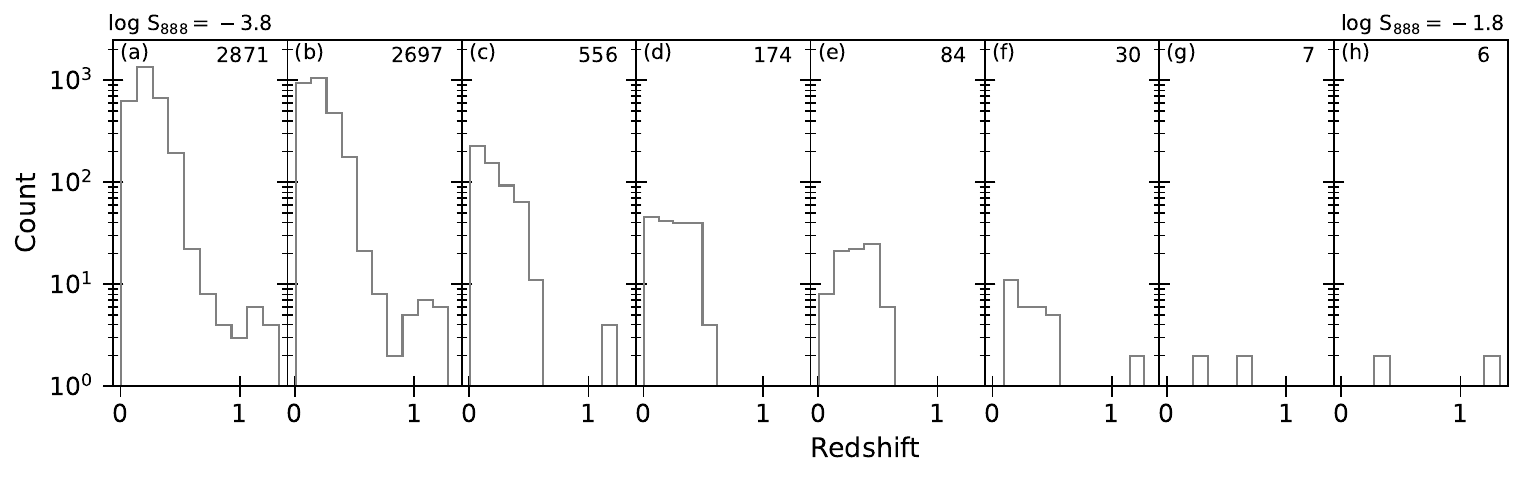}
    \caption{The redshift distribution of GEM\,I sources, used as the template for assigning redshifts to GEM\,III sources, arranged according to the bins in figure \ref{fig:S_1d}. The x-axis in each panel corresponds to redshift, and the y-axis shows the counts. The radio flux density increases along the x-axis. The number of sources is shown in the upper right corner of each panel.}
    \label{fig:1d_z_panel}
\end{figure*}

Figure \ref{fig:1d_z_panel} shows the redshift distribution in each bin of GEM\,I as seen in Figure \ref{fig:S_1d}. Here, the radio flux density increases along the $x$-axis. The bins corresponding to the lowest radio flux densities contain many sources with redshifts. We follow the same process here as for GEM\,II to assign redshifts. 
\begin{figure}[h!]
    \centering
    \includegraphics[width=\linewidth]{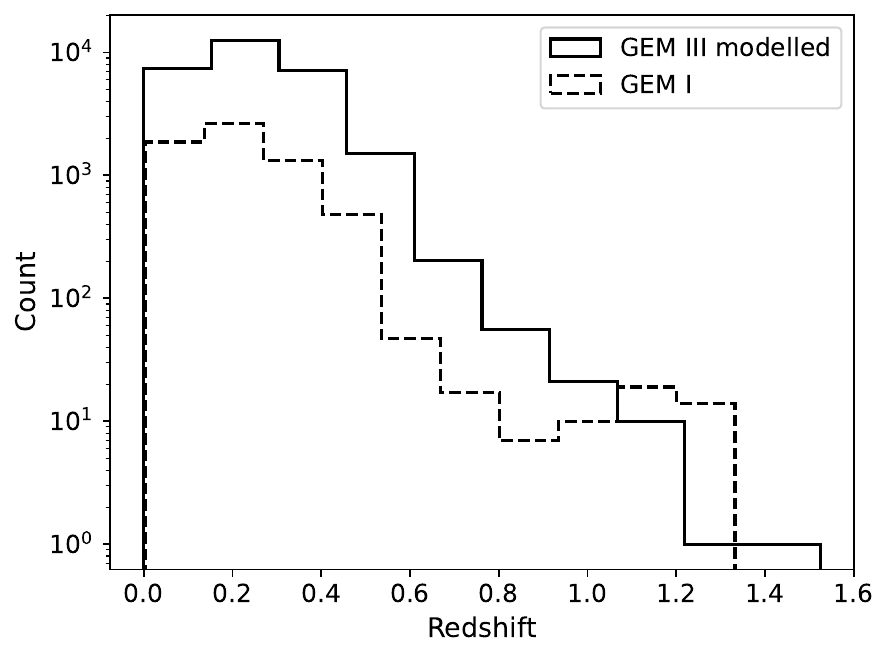}
    \caption{The modelled redshift distribution of GEM\,III sources (solid line). The redshift distribution of GEM\,I sources used for modelling is also shown (dashed line).}
    \label{fig:1d_z}
\end{figure}

Figure \ref{fig:1d_z} shows the distribution of modelled GEM\,III redshifts (solid histogram). It is evident that GEM\,II and GEM\,III sources have similar ranges of modelled redshift distribution. GEM\,II sources have the advantage of having $m_r$, putting an additional physical constraint on the estimation procedure. In the case of GEM\,III sources, however, only the radio flux densities constrain the possible redshifts assigned to the sources. The redshifts allotted for GEM\,III sources have a distribution that mimics the input GEM\,I sample, but as we later show, this is not likely to be accurate. We revisit this when deriving the RLFs in \S\,\ref{sec:rlf}. We now have the EMU sample with 39\,812 sources, all of which either have spectroscopic or modelled redshifts.
\section{AGN diagnostics}\label{sec:agn_diagnostics}
A plethora of techniques exist for performing AGN classification. Since the number of ionising photons from AGN and SFGs are different and they are distinct in their features across the electromagnetic spectrum \citep{1983ApJ...264..105F,1984A&AS...55...15S}, classification can be done using optical emission lines (e.g., \citealt{1981PASP...93....5B,1987ApJS...63..295V, 2011ApJ...736..104J}), photometric colours (optical; e.g., \citealt{2011ApJ...742...46T}, IR; e.g., \citealt{2012ApJ...753...30S}), or spectral energy distributions (e.g., \citealt{2022MNRAS.509.4940T,2024PASA...41...16P}). It is worth noting here that the AGN discussed here belong to the radiatively efficient type, omitting the low-excitation radio galaxies (LERGs) from this approach \citep{2012MNRAS.421.1569B,2017MNRAS.464.1306C,2024PASA...41...16P}.

Here we use the emission line classification proposed by \citet{1981PASP...93....5B}, the BPT diagram. Using the emission line ratios [N{\sc ii}]/H$\alpha$ and [O{\sc iii}]/H$\beta$, the BPT diagram classifies the galaxies based on their dominant ionising radiation source: star formation or black hole accretion. The classification scheme was extended by \citet{2001ApJS..132...37K} with a theoretical maximum on the starburst population and \citet{2003MNRAS.346.1055K} with a semi-empirical demarcation between SFGs and AGN. This revised scheme separates a given galaxy population into three: star forming, composites, and AGN. The emission lines used in the BPT diagram for classification have been shown to evolve as a function of redshift \citep{2013ApJ...774..100K}. We do not need to address this explicitly in our AGN classification, as we subsequently redefine our AGN criterion, as described below in \S\,\ref{sec:z_redist}. The sample used here does not have any S/N cut-off applied to the relevant spectral lines in order that the entire sample can be retained. Repetition of the analysis with an S/N>2 limit applied on the Balmer lines reduces our sample size by 30\% of the current size, but does not noticeably change our final derived RLFs. We choose the emission line based method for classification here, as it is the most common in other studies that estimate LFs \citep[e.g.,][]{2007MNRAS.375..931M,2016MNRAS.460....2P}. 
\begin{figure}[h!]
    \centering
    \includegraphics[width=\linewidth]{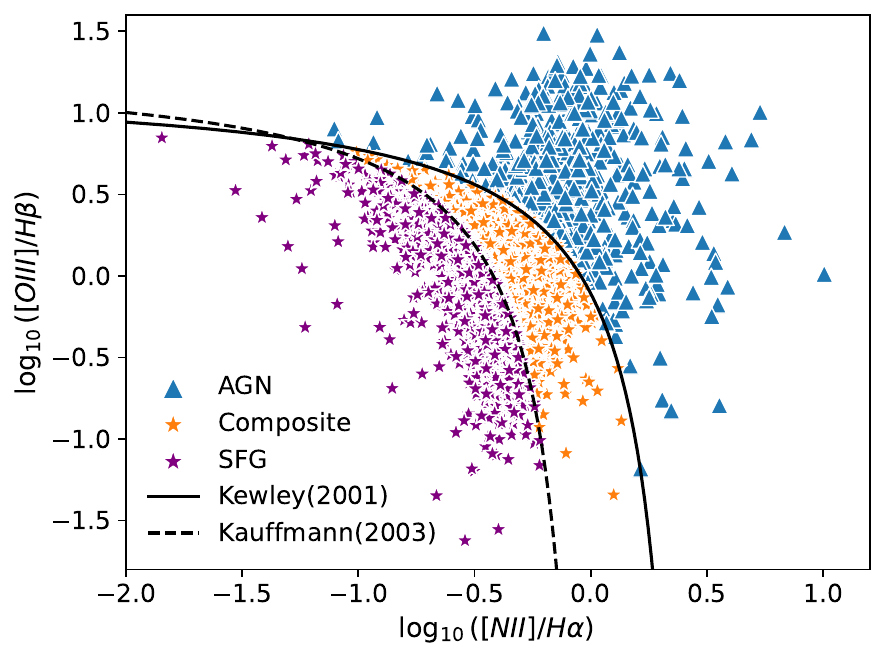}
    \caption{The BPT diagram: emission line classification scheme using the line ratios [NII/H$\alpha$] and [OIII]/H$\beta$. The diagram classifies GEM\,I sources into SFGs, composite galaxies, and AGN. The purple stars represent SFGs, the orange stars represent composite galaxies, and the blue triangles represent AGN. The black solid line corresponds to the \citet{2001ApJS..132...37K}, and the black dotted line corresponds to the \citet{2003MNRAS.346.1055K} separation.}
    \label{fig:bpt_z}
\end{figure}

GEM\,I sources, having optical spectra available, are classified using the BPT diagram in Figure \ref{fig:bpt_z}, where blue triangles represent AGN, orange stars represent composite galaxies, and purple stars represent SFGs. Of the 6425 GEM\,I sources, 4229 are SFGs, 1569 are composites, and the remaining 627 are AGN. Since spectral lines are required to perform this classification, we cannot apply the method directly to GEM\,II and GEM\,III. The choice of an emission line based classification, however, again constrains our ability to probe a larger dataset. To address this and assign classifications to the GEM\,II+GEM\,III samples, we again follow a statistical approach where classifications from GEM\,I sources are extended to both GEM\,II and GEM\,III.

GEM\,I sources are divided into bins of radio flux densities. Galaxies in each of these bins have a BPT classification. Composite galaxies are included among SFGs here because the primary study that we compare the results against \citep{2007MNRAS.375..931M} treats composites in this way \citep[see also,][]{2000PASA...17..234J,2002MNRAS.329..227S}. The combined GEM\,II+GEM\,III sample is also divided into the same bins of radio flux densities. Probability mass functions (PMFs) are constructed for each bin where the probabilities of AGN/SFG come from the fraction of AGN/SFG over the total number of sources in that bin. For instance, if a given bin contains 565 AGN and 5194 SFGs, the constructed PMF would correspond to $\sim$10\% AGN and $\sim$90\% SFG.

PMFs were constructed for the four bins considered (-3.89, -2.78, -1.67, -0.57, and 0.54 Jy, in log units)
, and the probabilities of AGN are 9.81\%, 11.7\%, 3.89\%, and 42.8\%, respectively. The bin sizes are chosen such that there are enough AGN/SFG classifications in each bin so that the newly assigned classifications are statistically robust. For a given bin, a number was chosen randomly according to the PMF of that bin to assign a classification to each source. The process was repeated, assigning a class to all sources in both GEM\,II and GEM\,III samples. This results in the entire EMU sample having a classification with 4047 AGN and $35\,765$ SFGs.
\section{The 888 MHz radio luminosity function}\label{sec:rlf}
\subsection{Sample properties}\label{sec:sample_properties}
The EMU sample now has radio sources with their redshifts and AGN classification.
\begin{figure*}[h!]
    \centering
    \includegraphics[width=0.9\linewidth]{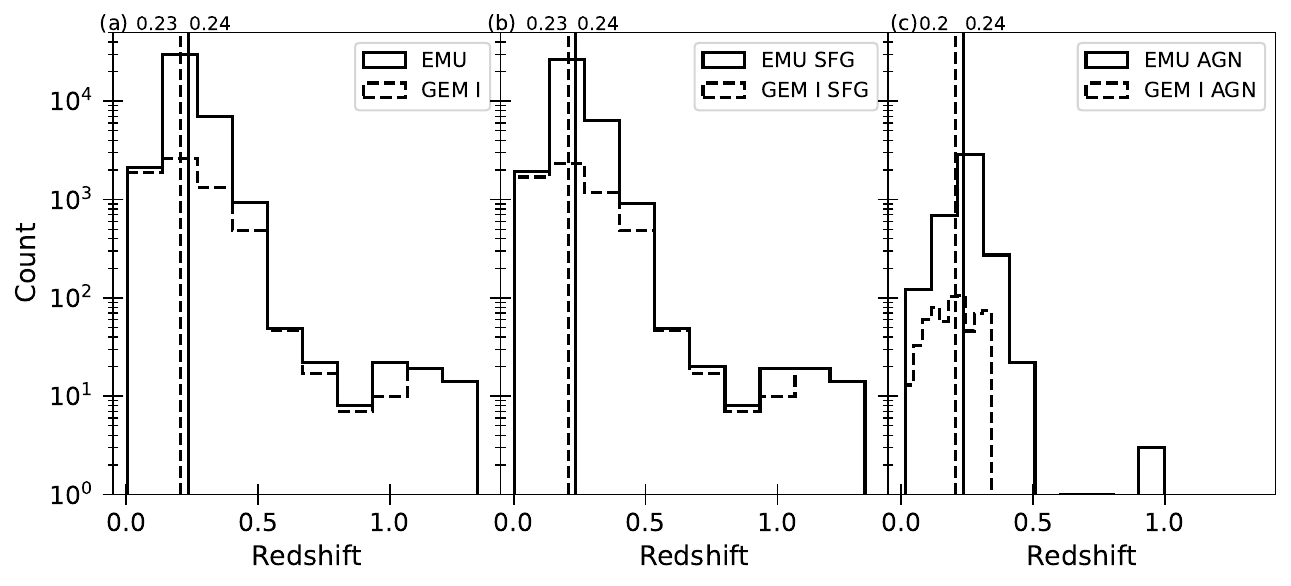}
    \caption{The redshift distributions of $(a)$ the EMU sample (solid line) and the GEM\,I sample (dashed line), $(b)$ EMU SFGs (solid line) and GEM\,I SFGs (dashed line), and $(c)$ EMU AGN (solid line) and GEM\,I AGN (dashed line), before reassigning redshifts (see 
    \S\,\ref{sec:z_redist}) and AGN classification (see 
    \S\,\ref{sec:z_redist}). The vertical dashed line in each panel corresponds to the median redshift of GEM\,I, and the vertical solid line shows the median redshift of the EMU sample. The median values of each class are shown on top of each panel.}
    \label{fig:emu_sfg_agn_z}
\end{figure*}
Figure \ref{fig:emu_sfg_agn_z}$a$ shows the redshift distribution of the EMU sample (solid histogram) and the GEM\,I sub-sample (dashed histogram). The solid and dashed vertical lines correspond to the median redshifts of the EMU and GEM\,I samples, respectively. Figures \ref{fig:emu_sfg_agn_z}$b$ and \ref{fig:emu_sfg_agn_z}$c$ show the same for star formers in EMU and GEM\,I  and AGN in EMU and GEM\,I, respectively. EMU AGN have similar median redshift (0.24), comparable to the EMU star formers (0.24). This is striking since the studies in the local universe report AGN with a higher median redshift than star formers \citep{2007MNRAS.375..931M,2016MNRAS.460....2P,2016MNRAS.457..730P}. This directly impacts the LFs derived below and is discussed in detail in \S\,\ref{sec:lf_results}.
\begin{figure*}[h!]
    \centering
    \includegraphics[width=0.9\linewidth]{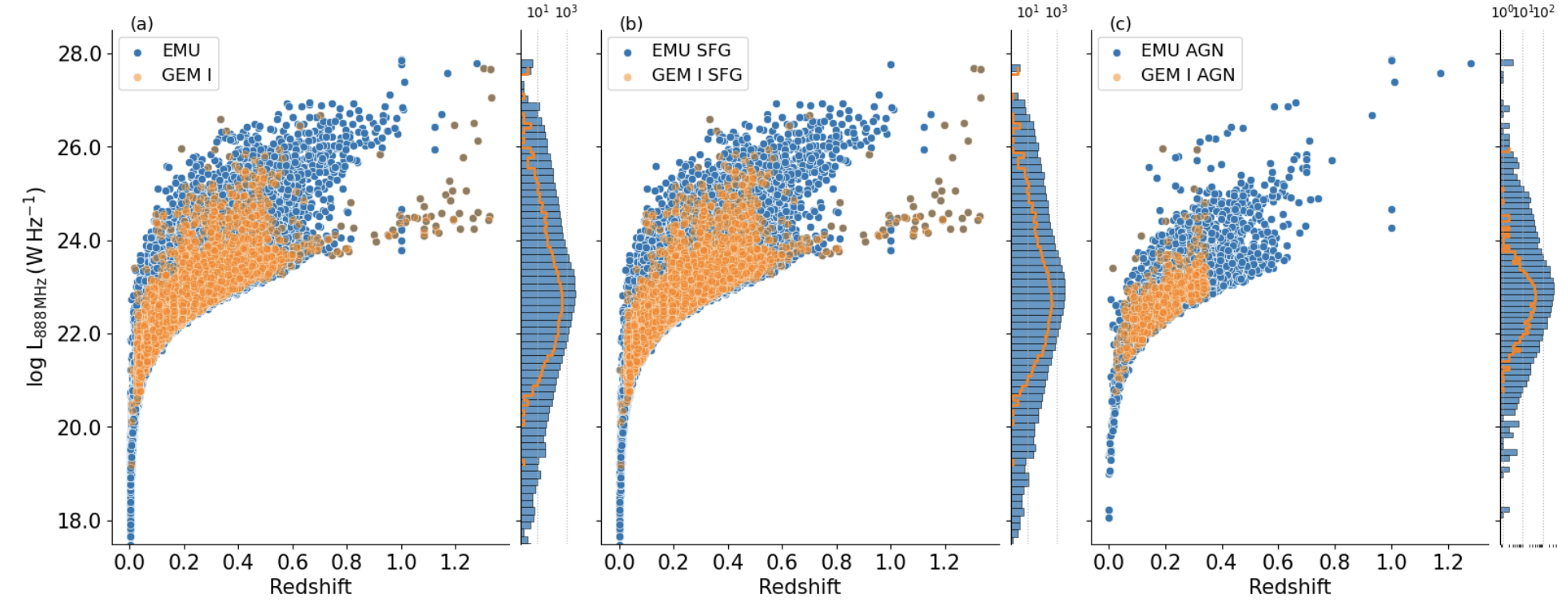}
    \caption{The figure shows the variation of the $\rm888\,MHz$ luminosities with redshift of $(a)$ the entire EMU sample (blue circles) and GEM\,I sample (orange circles), $(b)$ EMU star formers (blue circles) and GEM\,I star formers (orange circles), and $(c)$ EMU AGN (blue circles) and GEM\,I AGN (orange circles), before reassigning redshifts (see \S\,\ref{sec:z_redist}) and AGN classification (see \S\,\ref{sec:z_redist}). The marginal histogram on the y-axis shows the distribution of $\rm888\,MHz$ luminosities in $\log$ units with the same colours as the scatter plots.}
    \label{fig:lf_lum_z_dist}
\end{figure*}

Figure \ref{fig:lf_lum_z_dist} shows the $888\,\rm MHz$ radio luminosities as a function of redshift, where the EMU sample (blue circles) and the GEM\,I sample (orange circles) are shown in Figure \ref{fig:lf_lum_z_dist}$a$, the respective star formers in Figure \ref{fig:lf_lum_z_dist}$b$, and AGN in Figure \ref{fig:lf_lum_z_dist}$c$. The lower envelope results from the flux limit ($200\,\mu\rm Jy$) of the EMU early science survey \citep{2022MNRAS.512.6104G}. It is evident from the figures that radio luminosities of EMU AGN are distributed in the same range as that of EMU SFGs. Typically, AGN are the dominant radio sources above $1.4\,\rm GHz$ luminosities of $10^{23}\,\rm W\,Hz^{-1}$ \citep{1989ApJ...338...13C,1992ARA&A..30..575C,2016MNRAS.457..730P}. However, we see more star formers in the EMU sample in this regime, which we attribute to our statistical approach to redshift estimation. AGN that are actually at a higher redshift, if assigned a lower redshift value, will appear in the sample as a low luminosity source. We address this in \S\,\ref{sec:lf_results} when discussing the RLFs.
\subsection{Volume corrections: The classical $1/V_{max}$ method}\label{sec:v_max}
It is typical for flux-limited samples to become progressively more incomplete as the flux limit is approached. The completeness of a catalogue can be improved by adding missing objects from other sources \citep{1961MNRAS.122..263K} or each entry of the catalogue can be statistically weighted by the magnitude-dependent incompleteness function \citep{1979ApJ...232..352S}. Several methods exist, but we focus on the $1/V_{max}$ method. 

\citet{1968ApJ...151..393S} devised the $V/V_{max}$ technique to test and to correct for incompleteness. Here, $V$ is the sample volume between the galaxy and the observer, and $V_{max}$ represents the maximum volume in which the galaxy can exist without dropping below the flux limit (in this case, the radio flux density limit, $200\,\mu\,\rm Jy$ and the optical magnitude, $m_r=19.8$). \citet{1976ApJ...207..700F} showed that instead of having the number of galaxies in a given luminosity bin divided by $V_{max}$, the LF can be estimated by the sum, $\sum\,(1/V_{max})$ over all galaxies in that bin.

To calculate the LF, the EMU sample is divided into equally spaced redshift bins ($\Delta z=0.1)$. Each redshift bin is further divided into equally spaced luminosity bins ($\Delta\log L=0.25$). Construction of the LF requires, along with volume corrections, the flux-dependent incompleteness as discussed in \S\,\ref{sec:final_sample}, and the area correction, since the EMU early science observation covers a patch of the sky, the G23 field, totalling $83~\rm{deg^2}$ \citep{2022MNRAS.512.6104G}. Including all these contributions, the RLF is calculated as,
\begin{equation}
    \phi(L)=\frac{1}{\Delta\log L}\sum_{i=1}^N\,\frac{1}{V_{max,i}},
    \label{eq:radio_lf}
\end{equation}
where $i$ runs through each galaxy in a given redshift and luminosity bin. $V_{max,i}$ is calculated as
\begin{equation}
    V_{max,i}=\left[V(z_{max})-V(z_{min})\right]\,C_i,
    \label{eq:v_max}
\end{equation}
where $z_{min}$ is the lower bound of a given redshift bin and $z_{max}$ is the upper bound of that redshift bin, or it is the redshift at which the galaxy falls below the flux limit of the survey, whichever is the minimum. $C_i$ is the geometrical and statistical correction factor that takes into account the observed area and flux-dependent completeness:
\begin{equation}
    C_i=\frac{A}{41\,253\,\rm{deg^2}}\,C_{S_{888\,\rm{MHz}}},
    \label{eq:c(z)}
\end{equation}
where $A=83\,\rm{deg^2}$ and $C_{S_{888\,\rm{MHz}}}$ is the flux-dependent completeness factor (\S\,\ref{sec:final_sample}). Rest frame luminosities \citep{2002astro.ph.10394H} are calculated by assuming a $\Lambda$CDM cosmology following the prescription in \citet{2017A&A...602A...5N} for K-correction with $\alpha=-\,0.7$, where $\alpha$ is the radio spectral index (we adopt the convention,  $S_\nu\,\propto\,\nu^\alpha$).
\section{Initial outcomes}\label{sec:lf_results}
\begin{figure*}[h!]
    \centering
    \includegraphics[width=\linewidth]{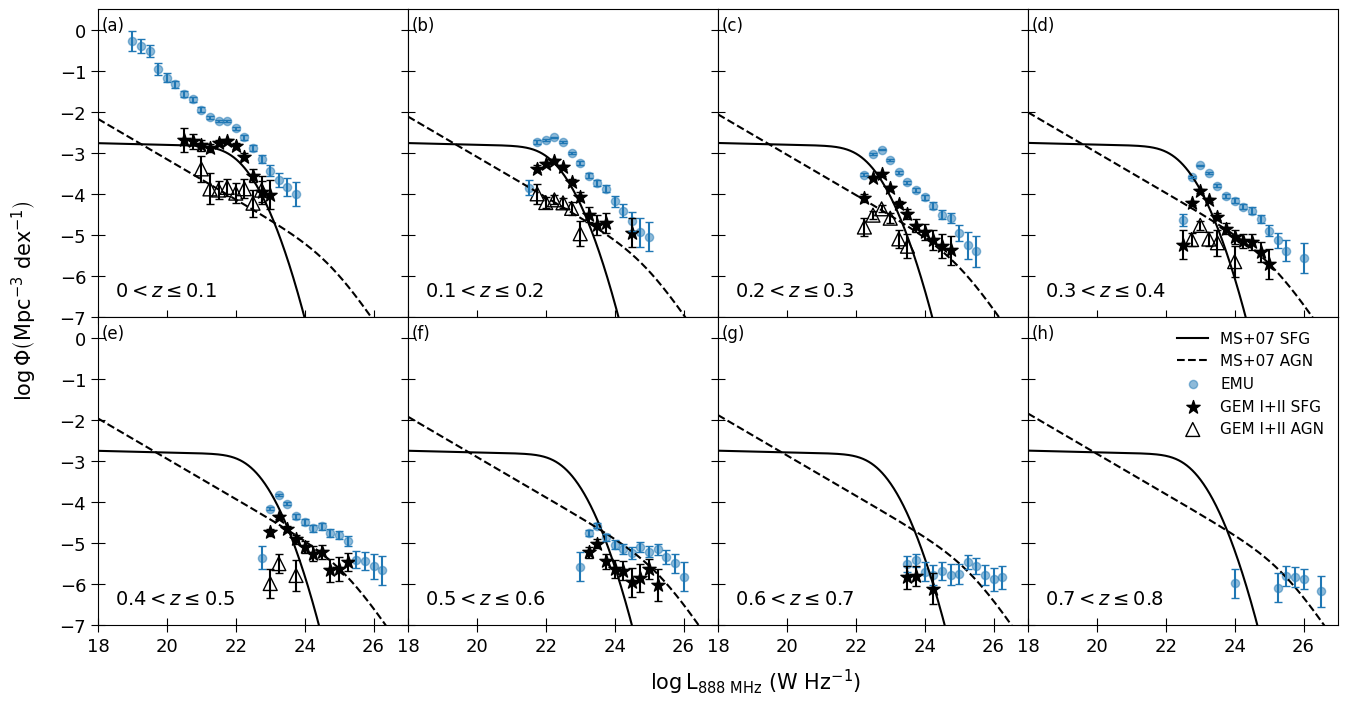}
    \caption{The $\rm 888\,MHz$ RLFs of the EMU sample (blue circles) compared with GEM\,I+II SFGs (the stars) and GEM\,I+II AGN (the triangles). The error bars assume that the numbers are Poisson distributed. The solid and dashed lines correspond to the parametric fits of \citet{2007MNRAS.375..931M} for SFGs and AGN, respectively, extrapolated to 888 MHz assuming $\alpha=-\,0.7$. The bins $a-h$ correspond to the RLFs at progressively increasing redshifts where the \citet{2007MNRAS.375..931M} lines are allowed to evolve in a pure luminosity fashion as described in the text. It is worth noting that how well the total GEM\,I+II RLF follow the \citet{2007MNRAS.375..931M} lines, up to $z=0.5$ (see text for details). It is also crucial to note that the redshifts of GEM\,II sources are modelled (close to 36\% of the GEM\,I+II sample). The low luminosity end starts to drop off as we move to higher redshifts, which is a consequence of the flux limit of the survey. It is also interesting to note that high redshift bins lack AGN, and there are star formers at high radio luminosities ($\rm >10^{24}\,W\,Hz^{-1}$) in these bins.}
    \label{fig:GEM_I_II_LF}
\end{figure*}

Figure \ref{fig:GEM_I_II_LF} shows our resulting $888\,\rm{MHz}$ RLFs in eight redshift bins. In each panel, the blue circles represent the EMU sample, the stars and triangles correspond to the SFGs and AGN, respectively, in the combined GEM I and GEM II (hereafter GEM I+II)  sample. The error bars are derived by assuming a Poisson distribution for the sources in each luminosity bin. The solid and dashed lines represent the parametric fits to the SFG and AGN RLFs, respectively,  derived by \citet{2007MNRAS.375..931M} (labelled as MS+07) using the analytical functions from \citet[][Eq. \ref{eq:saunders}]{1990MNRAS.242..318S} for SFGs and \citet{1990MNRAS.247...19D} and \citet{2001AJ....121.2381B} for AGN, converted from $1.4\,\rm{GHz}$ to $888\,\rm{MHz}$ assuming $\alpha=-\,0.7$. For higher redshifts, the fitted lines are allowed to evolve assuming PLE (Eq. \ref{eq:ple}), where $k_L$ is taken to be 3 \citep{2007MNRAS.381..211S,2011ApJ...740...20P,2013MNRAS.436.1084M,2016MNRAS.457..730P}, and the redshift is the mean of the bin. We do not fit our derived LFs with an analytical form; rather, we compare them with previous results.

It is clear in Figure \ref{fig:GEM_I_II_LF} that the RLF constructed using the modelled redshifts (the blue circles) results in an overestimate with respect to the \citet{2007MNRAS.375..931M} result. This discrepancy arises due to the use of the GAMA redshift distribution to sample radio source redshifts, since most are actually likely to be lying at higher redshifts.

GEM\,I+II sources (stars for SFGs and triangles for AGN) have redshifts that are both spectroscopic and modelled (from the estimation procedure described in \S\,\ref{sec:z_estimation}), with $\approx 36\%$ of them being modelled. It is remarkable to see that the total RLF for the GEM\,I+II sample is in reasonably good agreement with the \citet{2007MNRAS.375..931M} result up to a redshift of $z=0.5$. The LFs for the AGN are severely underestimated for $z\geq0.2$ (see panels c – h of Figure \ref{fig:GEM_I_II_LF}). Although, the GEM I+II SFGs and AGN show agreement with the respective \citet{2007MNRAS.375..931M} lines only upto $z=0.2$. This is most likely associated with the selection of only optically active AGN, as described in \S\,\ref{sec:agn_diagnostics}. We address this issue in \S\,\ref{sec:z_redist}. Since we are looking at a subsample that may not be complete, the SFGs and AGN in the high redshift bins are expected to fall in number, as can be seen in Figure \ref{fig:GEM_I_II_LF}. The GEM\,I+II star formers and AGN fall progressively below the faint end of the \citet{2007MNRAS.375..931M} lines as we move to higher redshifts, which is a consequence of the flux limit of the survey and the resulting incompleteness of the faint sources.

The mismatch between our GEM\,III LFs and the parametric fits by \citet{2007MNRAS.375..931M} can be attributed to the fact that a large portion of our modelled redshifts (close to 70\%) have the radio flux density as the sole constraint, most likely realistically lying at a larger redshift. We explore this further in the next section.
\subsection{Redistributing the redshifts and AGN classification}\label{sec:z_redist}
The GEM\,I+II sample, ignoring the AGN underestimate above $z=0.2$, is a good match for the \citet{2007MNRAS.375..931M} results at low redshifts ($0<z\leq0.5$) and is likely to accurately account for the entirety of that population given the high degree of spectroscopic completeness in the GAMA measurements. To confirm, we checked the agreement between the RLFs for GEM\,I and the \citet{2007MNRAS.375..931M} results and found that the total RLFs agree only upto $z=0.3$. This means, both GEM\,I and GEM\,II samples are required to accurately account for the RLFs up to $z=0.5$. To account for the overestimation in LF for radio sources without optical counterparts (GEM\,III, Figure \ref{fig:GEM_I_II_LF}), we adopt a different strategy.

\begin{figure}[h!]
    \centering
    \includegraphics[width=\linewidth]{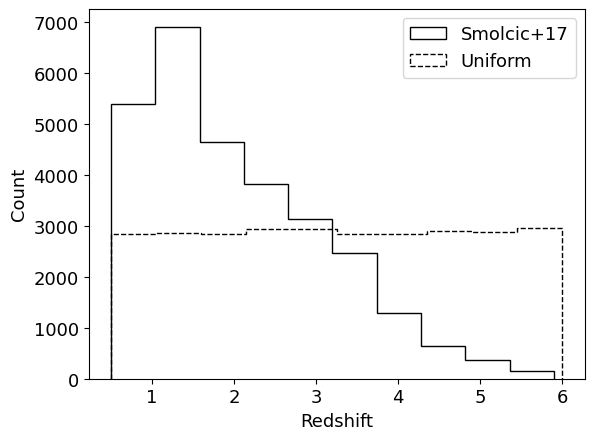}
    \caption{Solid line: the redshift distribution of the high redshift universe probed by \citet{2017A&A...602A...6S}. This scaled up version in the range $0.5<z<6$ is a more realistic representation that can be used to model the  GEM\,III source redshifts. Dashed line: the uniform redshift distribution sampling the high redshift Universe in the absence of a realistic high redshift distribution (\S\,\ref{sec:uniform_dist}).}
    \label{fig:SM_high_z}
\end{figure}

The GEM\,III sources are unlikely to be well represented by the GAMA redshift distribution ($z_{max}\approx0.5$). To account for this, we choose a more realistic high redshift distribution for these radio galaxies. \citet{2017A&A...602A...6S} derived the $\rm 1.4\,GHz$ RLF for AGN out to $z\approx5.5$ using the VLA-COSMOS observations and the COSMOS mutliwavelength data. We use this distribution and rederive the redshifts for the GEM\,III sources (as described in \S\,\ref{sec:emu_no_r-band}). Figure \ref{fig:SM_high_z} shows the \citet{2017A&A...602A...6S} distribution scaled up to match the GEM\,III sample size (solid line), along with a uniform distribution (dashed line) that we explore further in \S\,\ref{sec:uniform_dist} below.

The GEM\,III sources from the low redshift bins ($z\leq0.5$) have now been redistributed to redshifts above $z=0.5$ and the LFs are recalculated. The LFs for the AGN, however, are still severely underestimated for $z\geq0.3$. This is a consequence of neglecting radio loudness (or strong radio emission; $\rm L_{888\,MHz}\geq10^{ 23.5}\,W\,Hz^{-1}$) as a selection criterion for AGN. It has been established that luminous radio sources are typically AGN \citep{2000ApJS..126..133W}. We have around 10\% of the sources currently classified as star formers based solely on the BPT inspired classification, which also have high radio luminosities ($\rm L_{888\,MHz}\geq10^{ 23.5}\,W\,Hz^{-1}$) from the method above. Converting these radio luminosities to star formation rates \citep{2003ApJ...599..971H} would result in an SFR $\geq\rm\,100\,M_{\odot}\,yr^{-1}$, which is not typical for a large sample of SFGs. 
Erroneously classifying this population as AGN would have a negligible effect on the inferred luminosity functions. This approach has the inherent assumption that the GEM \,III sources are dominated by AGN. This argument is supported by the small fraction ($\approx0\%$) of SFGs with $\rm L_{1.4\,GHz}\geq10^{23.5}\,W\,Hz^{-1}$ \citep[see Figure 8 of][]{2007MNRAS.375..931M} and also that the GEM\,III sources are redistributed in to $z\geq0.5$.

To resolve this, we modify our classification of GEM\,II+GEM\,III to include radio luminosity as a criterion. The sources in the EMU sample with $\rm L_{888\,MHz}>10^{23.5}\,W\,Hz^{-1}$ are now labelled AGN, and the RLFs are recalculated. In addition, to understand the uncertainties in RLFs for different estimated redshifts, we also implement multiple realisations of the RLF. Table \ref{tab:radio_lf} contains the $888\,\rm MHz$ RLFs of the SFGs and AGN in the EMU sample.

\captionof{table}{The 888\,MHz RLFs ($\log\,\phi$) of the SFGs and AGN in the EMU sample and corresponding numbers (N) in each bin. The values here correspond to the logarithm of the median of the LFs calculated with one hundred estimated redshifts. The reported uncertainties are the $1\sigma$ deviations from the median (superscript) and Poisson errors (subscript). The luminosities shown here are the mean of each luminosity bin.}
\label{tab:radio_lf}
\vspace{3mm}
\begin{supertabular}{p{12mm}p{10mm}p{3mm}cp{3mm}c}\hline\hline
     {\small Redshift}&{\small $\rm\log L$}&\multicolumn{2}{c}{\small SFGs}&\multicolumn{2}{c}{\small AGN}  \\
     \cmidrule(lr){3-4}\cmidrule(lr){5-6}
     & & N & $\log\phi$ & N & $\log\phi$\\
    & ({\small$\rm W\,Hz^{-1}$}) & &({\small$\rm Mpc^{-3}\,dex^{-1}$})& &({\small$\rm Mpc^{-3}\,dex^{-1}$})\\
    \midrule
    \tablehead{%
    \midrule
    \hline
    {\small Redshift}&{\small $\rm\log L$}&\multicolumn{2}{c}{\small SFGs}&\multicolumn{2}{c}{\small AGN}  \\
     \cmidrule(lr){3-4}\cmidrule(lr){5-6}
     & & N & $\log\phi$ & N & $\log\phi$\\
    & ({\small$\rm W\,Hz^{-1}$}) & &({\small$\rm Mpc^{-3}\,dex^{-1}$})& &({\small$\rm Mpc^{-3}\,dex^{-1}$})\\
   \hline}
    $0.0$$<$$z$$\leq$$0.1$&\small  19.75& \small 3&\small$-1.59^{\pm 0.01}_{\pm 0.58}$&$-$&$-$\\[+0.5em] 
    &\small  20.0 &\small  6&\small $-1.95^{\pm 0.10}_{\pm 0.41}$&$-$&$-$\\[+0.5em] 
    &\small  20.25&\small  9&\small $-2.27^{\pm 0.11}_{\pm 0.33}$&$-$&$-$\\[+0.5em] 
    &\small  20.5 &\small  15&\small $-2.54^{\pm 0.09}_{\pm 0.26}$&$-$&$-$\\[+0.5em] 
    &\small  20.75 &\small 32 &\small $-2.60^{\pm 0.06}_{\pm 0.18}$&$-$&$-$\\[+0.5em] 
    &\small  21   &\small  64 &\small $-2.72^{\pm 0.04}_{\pm 0.13}$&\small  8&\small $-3.42^{\pm 0.04}_{\pm 0.35}$\\[+0.5em] 
    &\small  21.25&\small 113 &\small $-2.87^{\pm 0.03}_{\pm 0.09}$&\small 10&\small $-3.68^{\pm 0.05}_{\pm 0.32}$\\[+0.5em] 
    &\small  21.5 &\small 304 &\small $-2.76^{\pm 0.01}_{\pm 0.06}$&\small 25&\small $-3.83^{\pm 0.04}_{\pm 0.20}$\\[+0.5em] 
    &\small  21.75&\small 373 &\small $-2.68^{\pm 0.01}_{\pm 0.05}$&\small 22&\small $-3.89^{\pm 0.06}_{\pm 0.21}$\\[+0.5em] 
    &\small  22   &\small 247 &\small $-2.82^{\pm 0.01}_{\pm 0.06}$&\small 20&\small $-3.89^{\pm 0.05}_{\pm 0.22}$\\[+0.5em] 
    &\small  22.25&\small 122 &\small $-3.10^{\pm 0.01}_{\pm 0.09}$&\small 19&\small $-3.84^{\pm 0.05}_{\pm 0.23}$\\[+0.5em] 
    &\small  22.5 &\small  48 &\small $-3.49^{\pm 0.02}_{\pm 0.14}$&\small 12&\small $-4.04^{\pm 0.06}_{\pm 0.29}$\\[+0.5em] 
    &\small  22.75&\small  16 &\small $-3.92^{\pm 0.04}_{\pm 0.25}$&\small 11&\small $-3.96^{\pm 0.07}_{\pm 0.30}$\\[+0.5em] 
    &\small  23   &\small   8 &\small $-3.92^{\pm 0.04}_{\pm 0.35}$&$-$&$-$\\[+0.5em]

    \hline
    $0.1$$<$$z$$\leq$$0.2$
    &\small 21.75&\small 136 &\small $-3.37^{\pm 0.02}_{\pm 0.09}$&\small 22&\small $-4.00^{\pm 0.07}_{\pm 0.23}$\\[+0.5em] 
    &\small 22   &\small 498 &\small $-3.23^{\pm 0.02}_{\pm 0.05}$&\small 56&\small $-4.21^{\pm 0.03}_{\pm 0.13}$\\[+0.5em] 
    &\small 22.25&\small 786 &\small $-3.20^{\pm 0.01}_{\pm 0.04}$&\small 85&\small $-4.14^{\pm 0.02}_{\pm 0.11}$\\[+0.5em] 
    &\small 22.5 &\small 597 &\small $-3.29^{\pm 0.01}_{\pm 0.04}$&\small 78&\small $-4.17^{\pm 0.03}_{\pm 0.11}$\\[+0.5em] 
    &\small 22.75&\small 219 &\small $-3.68^{\pm 0.01}_{\pm 0.07}$&\small 51&\small $-4.29^{\pm 0.04}_{\pm 0.14}$\\[+0.5em] 
    &\small 23   &\small  67 &\small $-4.18^{\pm 0.02}_{\pm 0.12}$&\small 30&\small $-4.50^{\pm 0.06}_{\pm 0.18}$\\[+0.5em] 
    &\small 23.25&\small  25 &\small $-4.56^{\pm 0.03}_{\pm 0.20}$&\small 13&\small $-4.79^{\pm 0.08}_{\pm 0.28}$\\[+0.5em]
    &\small 23.5 &$-$ &$-$                                        &\small 22&\small $-4.62^{\pm 0.06}_{\pm 0.21}$\\[+0.5em]
    &\small 23.75&$-$ &$-$                                        &\small 15&\small $-4.77^{\pm 0.07}_{\pm 0.26}$\\[+0.5em] 
    &\small 24   &$-$ &$-$                                        &\small 11&\small $-4.90^{\pm 0.06}_{\pm 0.30}$\\[+0.5em]

    \hline
    $0.2$$<$$z$$\leq$$0.3$
    &\small 22.25&\small 90  &\small $-4.09^{\pm 0.06}_{\pm 0.11}$&\small 17&\small $-4.83^{\pm 0.04}_{\pm 0.24}$\\ [+0.5em]
    &\small 22.5 &\small 786 &\small $-3.61^{\pm 0.01}_{\pm 0.04}$&\small 85&\small $-4.54^{\pm 0.02}_{\pm 0.11}$\\ [+0.5em]
    &\small 22.75&\small 1010&\small $-3.48^{\pm 0.01}_{\pm 0.03}$&\small129&\small $-4.36^{\pm 0.02}_{\pm 0.09}$\\ [+0.5em]
    &\small 23   &\small 437 &\small $-3.80^{\pm 0.01}_{\pm 0.05}$&\small 80&\small $-4.52^{\pm 0.03}_{\pm 0.11}$\\[+0.5em]
    &\small 23.25&\small 145 &\small $-4.26^{\pm 0.02}_{\pm 0.08}$&\small 57&\small $-4.62^{\pm 0.05}_{\pm 0.13}$\\[+0.5em]
    &\small 23.5 &$-$&$-$                                         &\small 94&\small $-4.41^{\pm 0.02}_{\pm 0.10}$\\[+0.5em]
    &\small 23.75&$-$&$-$                                         &\small 49&\small $-4.68^{\pm 0.04}_{\pm 0.14}$\\ [+0.5em]
    &\small 24   &$-$&$-$                                         &\small 29&\small $-4.90^{\pm 0.06}_{\pm 0.19}$\\ [+0.5em]
    &\small 24.25&$-$&$-$                                         &\small 24&\small $-5.00^{\pm 0.05}_{\pm 0.20}$\\ [+0.5em]
    &\small 24.5 &$-$&$-$                                         &\small 16&\small $-5.12^{\pm 0.06}_{\pm 0.25}$\\ [+0.5em]
    &\small 24.75&$-$&$-$                                         &\small 10&\small $-5.27^{\pm 0.06}_{\pm 0.32}$\\[+0.5em]
    
    \hline
    $0.3$$<$$z$$\leq$$0.4$ 	
    &\small 22.75&\small 294&\small $-4.29^{\pm 0.02}_{\pm 0.06}$&\small 32&\small $-5.03^{\pm 0.12}_{\pm 0.18}$\\ [+0.5em]
    &\small 23   &\small 718&\small $-3.88^{\pm 0.01}_{\pm 0.04}$&\small 86&\small $-4.77^{\pm 0.03}_{\pm 0.11}$\\[+0.5em]
    &\small 23.25&\small 431&\small $-4.06^{\pm 0.02}_{\pm 0.05}$&\small 45&\small $-5.02^{\pm 0.05}_{\pm 0.15}$\\[+0.5em]
    &\small 23.5 &\small 64&\small $-4.84^{\pm 0.05}_{\pm 0.13}$&\small 110&\small $-4.62^{\pm 0.02}_{\pm 0.10}$\\[+0.5em]
    &\small 23.75&$-$&$-$                                       &\small 82 &\small $-4.72^{\pm 0.03}_{\pm 0.11}$\\ [+0.5em]
    &\small 24   &$-$&$-$                                       &\small 50 &\small $-4.90^{\pm 0.04}_{\pm 0.14}$\\ [+0.5em]
    &\small 24.25&$-$&$-$                                       &\small 36 &\small $-5.06^{\pm 0.04}_{\pm 0.14}$\\ [+0.5em]
    &\small 24.5 &$-$&$-$                                       &\small 32 &\small $-5.09^{\pm 0.04}_{\pm 0.18}$\\ [+0.5em]
    &\small 24.75&$-$&$-$                                       &\small 25 &\small $-5.22^{\pm 0.04}_{\pm 0.32}$\\ [+0.5em]
    &\small 25  &$-$&$-$                                        &\small 10 &\small $-5.56^{\pm 0.05}_{\pm 0.32}$\\[+0.5em]
    
    \hline
    $0.4$$<$$z$$\leq$$0.5$
    &\small 23   &\small 117&\small $-4.85^{\pm 0.03}_{\pm 0.09}$&\small 8 &\small $-5.66^{\pm 0.17}_{\pm 0.37}$\\[+0.5em]
    &\small 23.25&\small 384&\small $-4.33^{\pm 0.02}_{\pm 0.05}$&\small 24&\small $-5.49^{\pm 0.07}_{\pm 0.20}$\\[+0.5em]
    &\small 23.5 &\small 128&\small $-4.75^{\pm 0.03}_{\pm 0.09}$&\small 96&\small $-4.87^{\pm 0.02}_{\pm 0.10}$\\[+0.5em]
    &\small 23.75&\small 32&\small  $-5.29^{\pm 0.07}_{\pm 0.18}$&\small 68&\small $-4.99^{\pm 0.04}_{\pm 0.12}$\\ [+0.5em]
    &\small 24   &$-$&$-$                                        &\small 58&\small $-5.03^{\pm 0.03}_{\pm 0.13}$\\ [+0.5em]
    &\small 24.25&$-$&$-$                                        &\small 48&\small $-5.09^{\pm 0.04}_{\pm 0.14}$\\ [+0.5em]
    &\small 24.5 &$-$&$-$                                        &\small 37&\small $-5.19^{\pm 0.04}_{\pm 0.16}$\\ [+0.5em]
    &\small 24.75&$-$&$-$                                        &\small 26&\small $-5.35^{\pm 0.05}_{\pm 0.20}$\\ [+0.5em]
    &\small 25   &$-$&$-$                                        &\small 17&\small $-5.51^{\pm 0.07}_{\pm 0.24}$\\ [+0.5em]
    &\small 25.25&$-$&$-$                                        &\small 18&\small $-5.47^{\pm 0.06}_{\pm 0.24}$\\[+0.5em]
    
    \hline
    $0.5$$<$$z$$\leq$$0.6$
    &\small 23.25&\small 196 &\small $-4.77^{\pm 0.01}_{\pm 0.07}$&\small 25&\small $-5.58^{\pm 0.03}_{\pm 0.20}$\\[+0.5em]
    &\small 23.5 &\small 72  &\small $-5.18^{\pm 0.05}_{\pm 0.12}$&\small 304&\small $-4.54^{\pm 0.01}_{\pm 0.06}$\\[+0.5em]
    &\small 23.75&\small 33  &\small $-5.43^{\pm 0.07}_{\pm 0.17}$&\small 138&\small $-4.82^{\pm 0.01}_{\pm 0.09}$\\[+0.5em]
    &\small 24   &\small 6   &\small $-5.83^{\pm 0.07}_{\pm 0.41}$&\small 100&\small $-4.95^{\pm 0.01}_{\pm 0.10}$\\ [+0.5em]
    &\small 24.25&$-$&$-$                                         &\small 60 &\small $-5.12^{\pm 0.04}_{\pm 0.13}$\\ [+0.5em]
    &\small 24.5 &$-$&$-$                                         &\small 44 &\small $-5.28^{\pm 0.04}_{\pm 0.15}$\\ [+0.5em]
    &\small 24.75&$-$&$-$                                         &\small 35 &\small $-5.37^{\pm 0.04}_{\pm 0.17}$\\ [+0.5em]
    &\small 25   &$-$&$-$                                         &\small 20 &\small $-5.53^{\pm 0.06}_{\pm 0.22}$\\ [+0.5em]
    &\small 25.25&$-$&$-$                                         &\small 28 &\small $-5.46^{\pm 0.04}_{\pm 0.19}$\\ [+0.5em]
    &\small 25.5 &$-$&$-$                                         &\small 9 &\small $-5.78^{\pm 0.06}_{\pm 0.33}$\\[+0.5em]
    
    \hline
    $0.6$$<$$z$$\leq$$0.7$
    &\small 23.25&\small 10&\small $-6.16^{\pm 0.03}_{\pm 0.32}$&$-$&$-$\\[+0.5em]
    &\small 23.5 &$-$&$-$                           &\small 215&\small $-4.82^{\pm 0.01}_{\pm 0.07}$\\[+0.5em]
    &\small 23.75&$-$&$-$                           &\small 209&\small $-4.78^{\pm 0.01}_{\pm 0.07}$\\[+0.5em]
    &\small 24   &$-$&$-$                           &\small 101&\small $-5.06^{\pm 0.01}_{\pm 0.10}$\\ [+0.5em]
    &\small 24.25&$-$&$-$                           &\small 60 &\small $-5.26^{\pm 0.01}_{\pm 0.13}$\\ [+0.5em]
    &\small 24.5 &$-$&$-$                           &\small 38 &\small $-5.43^{\pm 0.03}_{\pm 0.16}$\\ [+0.5em]
    &\small 24.75&$-$&$-$                           &\small 30 &\small $-5.52^{\pm 0.03}_{\pm 0.18}$\\ [+0.5em]
    &\small 25   &$-$&$-$                           &\small 27 &\small $-5.55^{\pm 0.03}_{\pm 0.19}$\\ [+0.5em]
    &\small 25.25&$-$&$-$                           &\small 16 &\small $-5.73^{\pm 0.06}_{\pm 0.25}$\\ [+0.5em]
    &\small 25.5 &$-$&$-$                           &\small 16 &\small $-5.74^{\pm 0.05}_{\pm 0.25}$\\[+0.5em]
    &\small 25.75&$-$&$-$                           &\small 10 &\small $-5.92^{\pm 0.05}_{\pm 0.32}$\\[+0.5em]
    
    \hline
    $0.7$$<$$z$$\leq$$0.8$
    &\small 23.5 &$-$&$-$                           &\small 175&\small$-5.00^{\pm 0.01}_{\pm 0.08}$\\[+0.5em]
    &\small 23.75&$-$&$-$                           &\small 381&\small$-4.62^{\pm 0.01}_{\pm 0.05}$\\[+0.5em]
    &\small 24   &$-$&$-$                           &\small 201&\small$-4.85^{\pm 0.01}_{\pm 0.07}$\\ [+0.5em]
    &\small 24.25&$-$&$-$                           &\small 118&\small$-5.05^{\pm 0.01}_{\pm 0.09}$\\ [+0.5em]
    &\small 24.5 &$-$&$-$                           &\small 55 &\small$-5.35^{\pm 0.01}_{\pm 0.13}$\\ [+0.5em]
    &\small 24.75&$-$&$-$                           &\small 54 &\small$-5.33^{\pm 0.01}_{\pm 0.14}$\\ [+0.5em]
    &\small 25   &$-$&$-$                           &\small 23 &\small$-5.66^{\pm 0.03}_{\pm 0.21}$\\ [+0.5em]
    &\small 25.25&$-$&$-$                           &\small 28 &\small$-5.62^{\pm 0.02}_{\pm 0.19}$\\ [+0.5em]
    &\small 25.5 &$-$&$-$                           &\small 20 &\small$-5.73^{\pm 0.03}_{\pm 0.22}$\\[+0.5em]
    &\small 25.75&$-$&$-$                           &\small 13 &\small$-5.77^{\pm 0.06}_{\pm 0.28}$\\
        \bottomrule\midrule
\end{supertabular}
\vspace{3mm}
\begin{figure*}[h!]
    \centering
    \includegraphics[width=\linewidth]{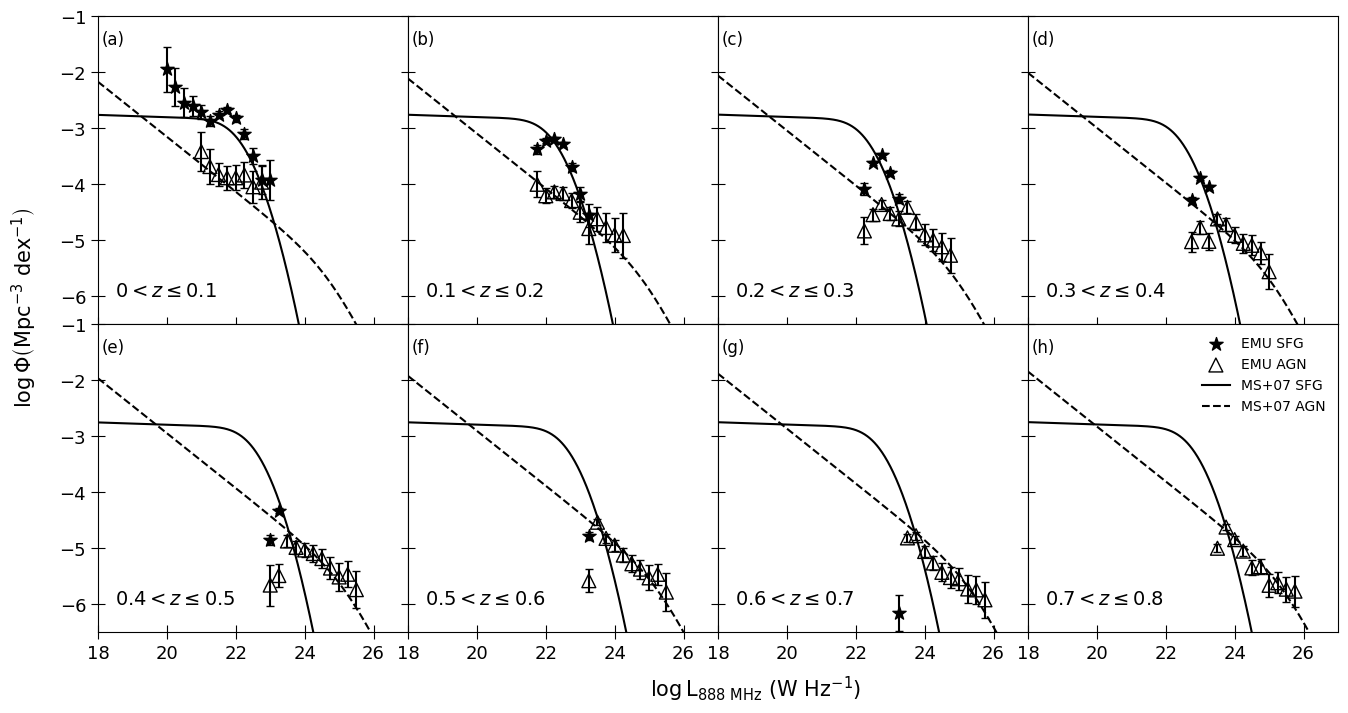}
    \caption{The $\rm 888\,MHz$ RLFs derived from the one hundred realisations. The symbols and lines follow the previous figures. The LFs plotted here are the median of the one hundred estimated LFs, and the error bars correspond to the Poisson errors. It is remarkable to see that the SFG and AGN LFs follow the \citet{2007MNRAS.375..931M} lines in every redshift bin. The upturn in the faint end of the low redshift LF, as discussed in the text, is also worth noting.}
    \label{fig:lf_median}
\end{figure*}

These LFs are estimated as the median of one hundred realisations, with redshifts assigned randomly, following the procedures detailed above each time. The uncertainties are calculated as the $1\sigma$ standard deviations of the one hundred realisations (superscript) and lie between 0.01 and 0.1. The actual errors are dominated by that from Poisson counting (subscript). We discuss these rederived LFs in the next section.
\section{Discussion}\label{sec:lf_discussion}
Figure \ref{fig:lf_median} shows the result of calculating one hundred realisations of the $\rm 888\,MHz$ RLFs with the error bars representing the Poisson errors. The symbols and lines follow Figure \ref{fig:GEM_I_II_LF}. These new RLFs exhibit significant improvement over Figure \ref{fig:GEM_I_II_LF} in how the star formers and AGN follow the \citet{2007MNRAS.375..931M} lines. Over the entire redshift range considered, with the applied PLE evolution, the star formers and AGN follow the reference LFs very well. 

Another interesting feature in Figure \ref{fig:lf_median} is the low-luminosity end of the lowest redshift bin as it rises considerably above the extrapolation from the \citet{2007MNRAS.375..931M} results. This is a consequence of the multiple realisations of redshifts sampling the low luminosity end, which are otherwise not likely to be detected. This is probably because of the sensitivity limits of the survey and the multiple realisations of the RLF most likely increases the likelihood of detection of these faint sources. Physically, this implies a larger number of faint radio sources at low redshifts, compared to the power-law tail of the Saunders function fit of \citet{2007MNRAS.375..931M}. The deviation begins close to $\rm L_{888\,MHz}\approx10^{20}\,W\,Hz^{-1}$ corresponding to an SFR of $\rm\approx0.07\,M_\odot\,yr^{-1}$ \citep{2003ApJ...599..971H}. Such an SFR value implies that the faint galaxies are of low stellar masses with $\rm\log\,M_*/M_\odot\leq\,9.5$ \citep{2007ApJ...660L..47N,2007ApJ...660L..43N} and they are potentially dwarf systems. Such upturns have also been seen in optical LFs estimated using GAMA data \citep[e.g.,][]{2012MNRAS.420.1239L,2014MNRAS.439.1245K,2020MNRAS.499..631V}.
The new generation of radio surveys probes large sky areas with high sensitivity, enabling the identification of numerous faint galaxies, and we expect that this feature may start to be detected as well at low radio luminosities.

Surprisingly, perhaps, random redshift assignments based on the radio flux densities and $m_r$ do a remarkable job in RLF estimation. This is in part a coincidence arising from the GAMA magnitude limit ($m_i<19.2$; \citealt{2011MNRAS.413..971D}), together with the high level of completeness of the GAMA survey. Consequently, radio sources with a GAMA counterpart will also be complete for the low redshift ranges ($z\leq 
0.5$) that GAMA is primarily sensitive to. Conversely, radio sources without a GAMA counterpart must lie at higher redshifts than those probed by GAMA. Some sources could be of low stellar mass ($\rm\log\,M_*/M_\odot\leq\,7.5-8$) and hence faint in radio luminosity, lying below the radio luminosities probed here.

Since the RLFs below $z\approx0.5$ match remarkably well with the \citet{2007MNRAS.375..931M} results, it is apparent that radio sources having optically faint counterparts ($m_r>19.8$) most likely lie at redshifts above $0.5$. Figure \ref{fig:lf_median} also implies that radio sources brighter than $\rm L_{888\,MHz}>10^{23.5}\,W\,Hz^{-1}$ are typically AGN.
\begin{figure*}[h!]
    \centering
    \includegraphics[width=\linewidth]{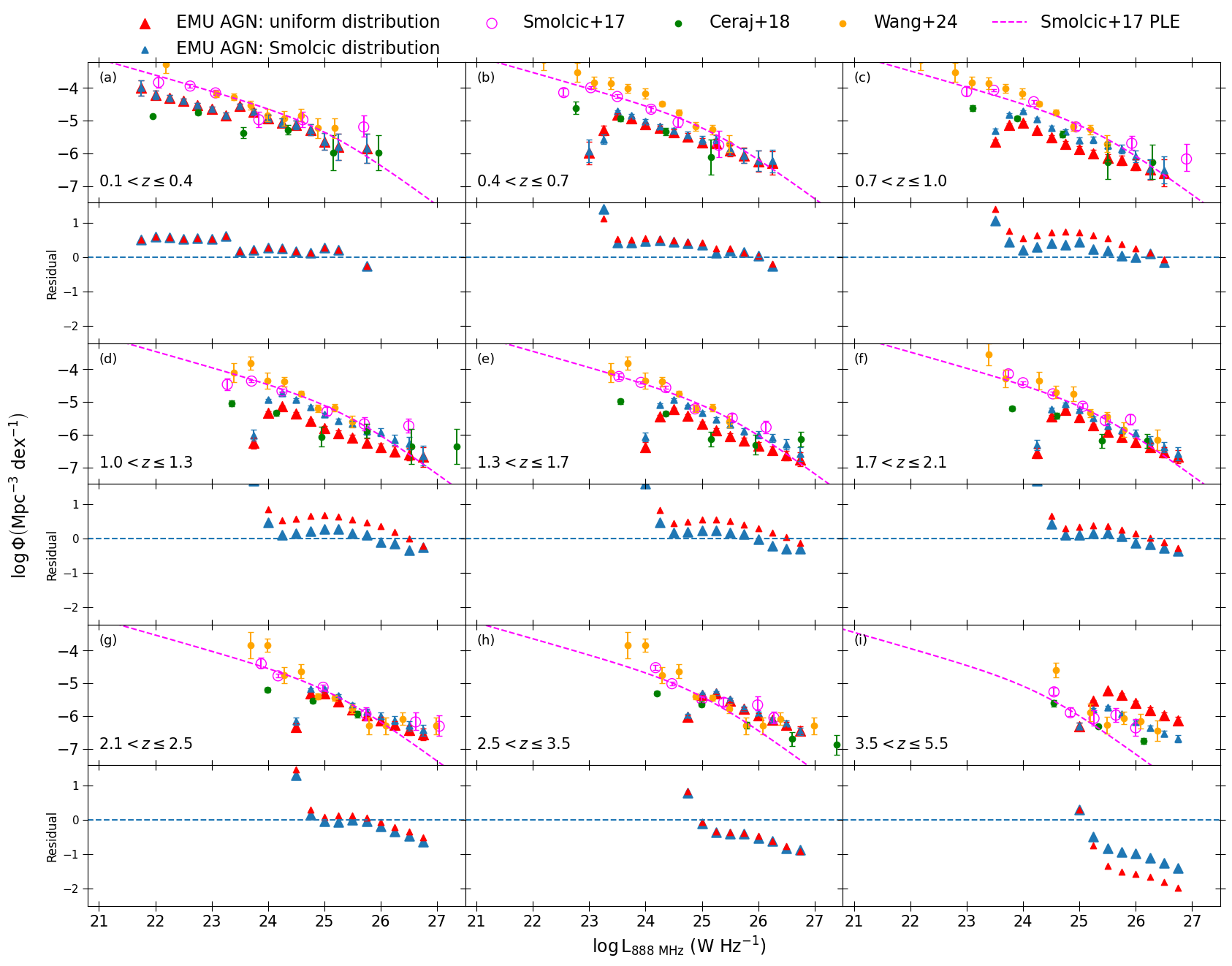}
    \caption{The 888 MHz RLFs derived using \citet{2017A&A...602A...6S} distribution (blue triangles), uniform distribution (red triangles), and 1.4 GHz RLFs derived by \citet{2017A&A...602A...6S} (magenta dots). The magenta dashed line shows the analytical fit to \citet{2017A&A...602A...6S} data which undergoes a pure luminosity evolution (PLE). The rows labelled 'Residual' show the difference between the \citet{2017A&A...602A...6S} fit to the LFs derived in this work (both red and blue points). Various results from the literature, converted from 1.4\,GHz to 888\,MHz assuming $\alpha=-\,0.7$ are also shown, as noted in the legend, with similar redshift ranges.}
    \label{fig:Uni_Sm}
\end{figure*}
\subsection{A scenario with no reference distribution}\label{sec:uniform_dist}
We now look at a scenario where we do not have a realistic high redshift distribution to sample redshifts from. To explore from a very basic level, we start with the simplest possible form: a uniform distribution in the range $0.5<z<6$. The redshifts for radio sources in GEM\,III are now sampled from this uniform distribution (Figure \ref{fig:SM_high_z}, dashed line) and the entire procedure is repeated to derive the new RLFs.

Figure \ref{fig:Uni_Sm} compares the RLFs derived using both the \citet{2017A&A...602A...6S} redshift distribution (blue triangles) and uniform redshift distribution (red triangles) with the radio AGN LFs of \citet{2017A&A...602A...6S} (magenta dots). The magenta dashed line represents the analytical fit derived by \citet{2017A&A...602A...6S} with an incorporated PLE. These RLFs are also compared against a bunch of recent works \citep[][]{2018A&A...620A.192C,2019A&A...625A.111B,2019ApJ...872..148C,2024A&A...685A..79W}, all measured at 1.4\,GHz and converted to 888\,MHz assuming $\alpha=-\,0.7$, as shown in the figure. The redshift bins in this plot are chosen to match \citet{2017A&A...602A...6S}. The residual plots quantify the logarithmic separation between the derived LFs (following the same colours as the LFs) with the \citet{2017A&A...602A...6S} analytical fit.

The RLFs we derived using the Smol\v{c}i\'c distribution (blue triangles) and the \citet{2017A&A...602A...6S} LFs (magenta circles) follow each other well in most of the redshift ranges probed here. In addition, our derived LFs are more densely sampled and extend to the high luminosity end (close to 1 dex brighter) towards progressively higher redshifts.

The LFs derived in this work using the Smol\v{c}i\'c (blue triangles) and the uniform redshift distribution (red triangles) agree with each other well in most redshift ranges. These LFs exhibit a drop in the lowest few luminosity bins which can be attributed to various reasons that still persist even after the completeness corrections, including the sensitivity limits of the surveys used. The separation between the two approaches a maximum of $\approx0.6$ dex in the highest redshift bin (at $\rm L_{888\,MHz}=10^{26}\,W\,Hz^{-1}$). These results point to two realisations, $(i)$ our statistical approach to measuring RLFs does surprisingly well even at $z\approx5.5$, and $(ii)$ even adopting a uniform distribution for radio galaxy redshifts up to $z=6$ leads to results broadly comparable, at levels notably better than an order of magnitude, in deriving the LFs. This agreement is also reflected in their respective residuals.

The RLF is a statistical description of how common radio galaxies of various luminosities are in the Universe. The lack of redshifts for a large fraction of the sample requires us to look for alternatives in calculating the RLFs. Our approach is the application of a simple statistical method to measure the RLF in the absence of redshfits. Rather than the properties of a single radio galaxy, we focus on the characteristics of a population and are successful in generating LFs that compare well with previous studies. The application of the uniform distribution demonstrates another viable option in measuring the RLFs in the absence of any additional multiwavelength information and a robustly constrained redshift distribution.
\section{Conclusion}\label{sec:conclusion}
We have used a sample of 39\,812 radio galaxies from the EMU early science observations to derive their RLFs at 888 MHz. Since only $\approx16\%$ of the radio galaxies have optical counterparts and hence redshifts, we modelled the redshift distributions of the radio sources using simple statistical tools, which is then used to derive RLFs. Since the redshift distribution of GAMA sources was not a realistic representation of high redshift radio galaxies, we used the \citet{2017A&A...602A...6S} redshift distribution to derive the RLFs of those high redshift radio galaxies. Considering a scenario where the high redshift radio sources do not have a multiwavelength counterpart, we assumed a uniform distribution for the high redshift radio galaxy redshifts and rederived the RLFs. The main conclusions are the following:
\begin{itemize}
    \item In the absence of observational counterparts and hence individual redshifts, modelling a redshift distribution for the entire sample has turned out to be a surprisingly effective tool in calculating the LFs.
    \item A probabilistic approach can help detect population that are less likely to be identified observationally. The faint end population seen in Figure \ref{fig:lf_median} serves as an example.
    \item Given a sparsely sampled dataset spanning any redshift range, our method can be used to create redshift distributions, hence densely populated samples. The RLFs derived using this distribution probes the high luminosity end, especially at higher redshifts. This feature is evident in Figure \ref{fig:Uni_Sm} while comparing the blue and magenta data points.
\end{itemize}

In the absence of accurate redshift information, it is clearly still possible to reproduce reliable properties for broad galaxy population statistics, using this kind of method. This has the advantage of leveraging large numbers of sources to improve the sampling of such distributions compared to small samples with precise redshift measurements. The limitations, of course, are that rare populations and accurate redshifts are not able to be reproduced.

In a follow up work we plan to use the LFs derived in this way to explore the star formation rate and AGN luminosity density evolution, drawing on the large numbers available to investigate dependencies on other galaxy properties, such as stellar mass.
\begin{acknowledgement}
This scientific work uses data obtained from Inyarrimanha Ilgari Bundara / the Murchison Radio-astronomy Observatory. We acknowledge the Wajarri Yamaji People as the Traditional Owners and native title holders of the Observatory site. The Australian SKA Pathfinder is part of the Australia Telescope National Facility (\url{https://ror.org/05qajvd42}) which is managed by CSIRO. Operation of ASKAP is funded by the Australian Government with support from the National Collaborative Research Infrastructure Strategy. ASKAP uses the resources of the Pawsey Supercomputing Centre. Establishment of ASKAP, the Murchison Radio-astronomy Observatory and the Pawsey Supercomputing Centre are initiatives of the Australian Government, with support from the Government of Western Australia and the Science and Industry Endowment Fund. This paper includes archived data obtained through the CSIRO ASKAP Science Data Archive, CASDA (\url{http://data.csiro.au}).

GAMA is a joint European-Australasian project based around a spectroscopic campaign using the Anglo-Australian Telescope. The GAMA input catalogue is based on data taken from the Sloan Digital Sky Survey and the UKIRT Infrared Deep Sky Survey. Complementary imaging of the GAMA regions is being obtained by a number of independent survey programmes including {\em GALEX\/} MIS, VST KiDS, VISTA VIKING, {\em WISE\/}, {\em Herschel}-ATLAS, GMRT and ASKAP providing UV to radio coverage. GAMA is funded by the STFC (UK), the ARC (Australia), the AAO, and the participating institutions. The GAMA website is \url{http://www.gama-survey.org/}.
\end{acknowledgement}


\bibliography{references}


\end{document}